\documentclass{amsart}
\usepackage{colortbl}

\newtheorem{thm}{Theorem}[section]

\newtheorem{rem}{Remark}[section]

\begin{document}
%---------------------------------------------------------------
\title{Weakly nonlocal irreversible thermodynamics}
\author{P. V\'an}
\address{
Budapest University of Technology and Economics\\
Department of Chemical Physics\\
1521 Budapest, Budafoki \'ut 8.} \email{vpet@phyndi.fke.bme.hu}

% \date{\today}
%---------------------------------------------------------------

\begin{abstract}
Weakly nonlocal thermodynamic theories are critically revisited. A
relocalized, irreversible thermodynamic theory of nonlocal phenomena is
given, based on a modified form of the entropy current and new kind of
internal variables, the so called {\em current multipliers}. The
treatment is restricted to deal with nonlocality connected to dynamic
thermodynamic variables. Several classical equations are derived,
including Guyer-Krumhansl, Ginzburg-Landau and Cahn-Hilliard type
equations.
\end{abstract}

\dedicatory{Dedicated to the memory prof. Istv\'an Gyarmati.}
\maketitle
%----------------------------------------------------------------

\section{Introduction}

In the last decades there has been a continuous interest in developing
generalized classical continuum theories to include memory and nonlocal
effects. By {\em memory effects} one means non instantaneous
interactions and actions in the continuum, that is the dynamics of the
continuum is determined not only by the present state but also by the
previous ones, the history of the continuum. By {\em nonlocal effects}
one means that the interaction between representative volume elements
is not local, the dynamics is determined not only by the closest
neighbours, the influence of farther volume elements is not negligible.
There are at least three basic methods to consider memory and nonlocal
effects in the dynamic equations. One can construct {\it hypercontinuum
theories} introducing new variables beside the space-time variables
(see e.g. \cite{Rog79a,Eri99b}). In case of {\it strongly nonlocal}
theories the constitutive functions are given by a space integral form
directly calculating the effects of the farther neighborhood (see e.g.
\cite{Ede74a,Rog82c}). The most popular are the so called {\it weakly
nonlocal} theories, where higher order space derivatives are introduced
into the governing equations than is usual in classical approaches. The
original governing equations, that one wants to generalize, can be the
balance equations of classical continuum physics or some other
equations, too. The choice of the method is more or less a matter of
convenience and experience, but without doubt the last method, the
construction of weakly nonlocal continuum theories is the most
attractive and popular among them \cite{Mau79a}.

The popularity of weakly nonlocal theories is partially based on
the convenience and the familiar knowledge of differential
equations. The other approaches are somehow too general to be
easily understandable and manageable, and bear some basic
controversies. First of all, in hypercontinua or in the strongly
nonlocal approach some requirements of a physical theory (e.g.
objectivity, distinction between material and structural
components) are not included and exploited. Moreover, the nonlocal
effects (e.g. a kind of weakening of interactions) are to be
connected to the Second Law at least in a general heuristical
sense. Seemingly there is no clear conceptual prescription in the
different formulations of the Second Law that would give a firm
method in hypercontinua or strongly nonlocal theories. On the
other hand, weakly nonlocal theories have an approximative
character. First we can consider derivatives with one order higher
than in the traditional equations and we can check the effects
experimentally and theoretically. More systematic considerations
should give a series of weakly nonlocal equations as a kind of
successive approximation.

It is interesting and instructive to review briefly the structure of
the theories describing memory effects and compare them with nonlocal
theories. The general memory functional method of rational
thermodynamics ({\em strong history dependence}) (see e.g. in
\cite{TruNol65b}) is the counterpart of strongly nonlocal theories. The
time related counterparts of weakly nonlocal theories ({\it weak
history dependence}) are the models that use rate dependent
constitutive relations in the basic balance laws. Both are well treated
and used approaches in this respect. However, the last development
(more than thirty years ago ) was the introduction of internal
variables of different kind, and in this way one can understand that
the balance structure and the Second Law can generate the rate
dependence (see e.g. in \cite{Ver97b}). This method preserves the clear
physical structure of classical irreversible thermodynamics and gives a
considerable flexibility in modelling physical phenomena. Let us
recognize that using internal variables instead of time derivatives in
the constitutive relations, the constitutive relations themselves are
strictly instantaneous, they depend only on state variables of the
material, therefore only on the present state. The internal variables
are related to the instantaneous state of the material, they are
'local' (in time), therefore, with their help the memory effects are
modelled with a local theory, we can say that they are {\em
relocalized}. The next table summarizes the situation:

\vskip 0.1in \setlength{\extrarowheight}{2mm}
\begin{tabular}{|l|p{1.5in}|p{1.5in}|}\hline
    & {Space}& {Time} \\ \hline\hline
Strongly nonlocal & space integrals & memory functionals \\\hline
Weakly nonlocal & gradient dependent constitutive functions
    & rate dependent constitutive relations \\\hline
Relocalized &\centering{???} & {internal variables} \\\hline
\end{tabular}
\vskip 0.1in

In this paper we investigate the possibility to fill the place of the
question marks in the table, to relocalize weak nonlocalities by
introducing new kind of internal variables in a way that is compatible
with the Second Law and with traditional methods of irreversible
thermodynamics. We can accomplish our task when the nonlocality is
connected to the so called dynamic internal variables (dynamic degrees
of freedom), those that vanish in (local) thermodynamic equilibrium.

The general strategy of weakly nonlocal approaches is to include space
derivatives, gradients of field quantities in the different classical
constitutive functions (therefore sometimes they are called 'gradient
theories'). The justification of these generalizations is based on
either checking the consequences and/or checking the requirements of
basic physical principles as, for example, different forms of
objectivity or/and the Second Law. Several approaches are used in
mechanics e.g. different gradient theories
\cite{KosWoj95a,Kos96p,CimKos97a,Val96a,Val98a,VarAif94a,Bed00a}, the
virtual power considerations \cite{Mau90a1,Mau80a} and the multifield
theories \cite{CapMar01a,Mar02a}. Instead of detailed reviewing and
criticising the different weakly nonlocal approaches our general remark
is, that they usually introduce new and disputable concepts that seem
to be too special to serve as foundation of a general nonlocal
thermodynamic theory. Moreover, they are not constructed as a general
approach to deal with nonlocal extensions of the transport equations of
classical irreversible thermodynamics, they usually give methods for
special fields of continuum physics (heat conduction, thermomechanics,
etc..).

\subsection{Requirements of the Second Law}

Introducing gradient dependence in an ad-hoc way one can get equations
with solutions contradicting with the Second Law. Even if we keep the
equations we should exclude some initial conditions and solutions
because of physical reasons (e.g. because they contradict the
experiments). We should and can avoid these difficulties applying a
systematic approach, a constitutive theory. To formulate a constitutive
theory for nonlocal effects we strongly need a clear formulation of the
Second Law and a construction method to get equations that
automatically fulfill this requirement. The fundamental method of
continuum physics to handle the situation is to separate the material
and the structural properties of the phenomena: one should use the
basic balance equations of the considered material and introduce
constitutive functions that correspond to the Second Law. The question
arises: what kind of Second Law, which formulation is the most
fruitful? In this paper we accept that there is an extensive entropy
function $s$, that should have a non negative production for supply
free processes, in pure dissipative systems. We will see that this
conservative standpoint is sufficiently general for our purposes. We do
not need to weaken the well founded theoretical basis of irreversible
thermodynamics and to exploit a freedom of any more general approach,
e.g. to introduce a nonextensive entropy function.

In non-equilibrium thermodynamics there are three traditional methods
to check the compatibility of constitutive function with the inequality
of the Second Law.
\begin{itemize}
\item {\em Irreversible thermodynamics} derives an expression of the
entropy production and introduces thermodynamic forces and currents
with linear or nonlinear relations between them
\cite{GroMaz62b,Gya70b}.
\item {\em Coleman-Noll procedure} investigates more systematically
the linear independence of the different possible forces and
currents \cite{ColNol63a}.
\item {\em Liu procedure} looks for the form of the constitutive
functions that ensures the validity of the entropy inequality with
any solutions of the governing differential equations, without
substituting them into the entropy balance \cite{Liu72a}.
\end{itemize}

The methods are ordered with growing exactness. The first method is
constructive. One can find a more detailed treatment regarding the
different formulations and exploitations of the Second Law inequality
for example in \cite{Hut77a,Sil97b,MusEhr96a}. In the following
considerations one of our basic aims is to combine the heuristic power
of the traditional force-current systems with a mathematically clearer
treatment.

%Here we only remark that contrary to other formulations our
%consideration preserves the stability structure of thermodynamics
%therefore it seems to be the most plausible from a physical point
%of view.

\subsection{Discrete and continuum}

Beyond the compatibility with the Second Law in all continuum
thermodynamic theories is important and instructive to clarify the
relation between discrete and continuum descriptions of physical
systems. Some continuum presentations pronounce to be self standing
continuum theories without any connection to thermodynamics of
homogeneous, discrete systems. Homogeneous bodies are considered as
special cases of continua. However, the relation is far more difficult
if we consider processes in homogeneous thermodynamic systems as it was
shown in \cite{Mat92a1}. Anyway, a conceptual classification gives the
basic heuristics. The relation of continuum and discrete systems is
based on {\em continuization principles}. Classical irreversible
thermodynamics is based on the {\em principle of local equilibrium}
\cite{GroMaz62b,GlaPri71b}, i.e. in that theory we suppose that
\begin{itemize}\item
physical quantities defined in ordinary (equilibrium)
thermodynamics of discrete systems and the relations between them
are valid in a continuum theory as well.
\end{itemize}

Particularly it means that in theories using the principle of
local equilibrium the entropy function depends only on equilibrium
state variables and the entropy current is classical.

In an internal variable theory we use the {\em principle of local
state} of Kestin \cite{Kes90a1,Kes93a1}, implicitly or explicitly
supposing that

\begin{itemize}\item
physical quantities defined in non-equilibrium thermodynamics of
discrete systems and the relations between them are valid in a
continuum theory, too.
\end{itemize}

Particularly it means that in theories using the principle of
local state the entropy function can depend on internal variables
beyond the equilibrium state variables and the entropy current is
classical. In this way we can get equations incorporating higher
order time derivatives and we can preserve the stability structure
of thermodynamics.

Let us remark that in the literature both concepts appear in a
controversial treatment (including the mentioned references). For
example de Groot and Mazur formulates the concept of "local
equilibrium" as follows:

"It will be now assumed that, although the total system is not in
equilibrium, there exists within small mass elements a state of "local"
equilibrium, for which the local entropy $s$ is the same function
previously defined (for equilibrium systems) of $u$, $v$ and $c_k$ as
in real equilibrium." \cite{GroMaz62b}

The controversy is contained in the concept of equilibrium. A correct
formulation and understanding incorporates a suitable nonequilibrium
theory of discrete (homogeneous) thermodynamic bodies including
internal variables started by Onsager and Machlup
\cite{OnsMac53a,MacOns53a} and continued by others
\cite{Mat92a1,Mat95a,Van95a,Mat96a1,MusDom96a,Mat96a2,Mat00a}, because
in the above statement an equilibrium system is necessarily time
dependent.

In nonlocal theories we are beyond the validity of local equilibrium
and local state and we deal with continuum systems that do not have a
discrete counterpart. In the following we introduce new kind of
internal variables to characterize the nonlocality of the continuum.
The nonlocal internal variables are current-like in some sense, and the
extension of the state space appears not only through the modified
entropy function but also through a modified entropy current. The local
character of the theory is preserved because the entropy function and
the entropy current function are local, do not depend on space
derivatives.

The necessity to use a generalized entropy current was pointed out
and emphasized several times by M\"uller \cite{Mul68a,MulRug98b}.
Regarding the internal variables it was Verh\'as who developed
M\"uller's generalization and introduced higher order corrections
with the help of transport of internal variables \cite{Ver83a}.
The importance of the entropy current describing nonlocal effects
was pointed out by several previously mentioned authors
\cite{MulRug98b,LebAta97a,MarAug98a}.

Our approach will be called {\em weakly nonlocal irreversible
thermodynamics} because we give conditions to transform the entropy
inequality into a solvable, quadratic form. In this way the
constitutive theory can fully preserve the predictive character of
irreversible thermodynamics. In this case the constitutive theory is
particularly easy, one can identify force-current systems with
classical, well established methods \cite{GroMaz62b,Gya70b}. The
applicability of the approach is demonstrated by deriving several
particular classical examples like weakly nonlocal heat conduction,
generalized Ginzburg-Landau and Cahn-Hilliard type equations.

In the next section we use Liu procedure with a precise prescription of
the state space, that is free from several weaknesses of the
traditional treatment. In this way the background and the validity of
our approach is clear and we will see among others why and in what
sense Ginzburg-Landau-like and Cahn-Hilliard-like equations play a
distinguished role in weakly nonlocal irreversible thermodynamics. In
the third section we treat more examples and some other problems by a
more heuristic approach to show the easy applicability. The last
section contains a summary and discussion.

\section{Weakly nonlocal irreversible thermodynamics}

There are some basic problems in the heuristic approach of
irreversible thermodynamics that should be addressed in a
consistent treatment:
\begin{itemize}
\item One should decide what are the constitutive functions and
what are the variables (derivatives) they are depending on. That
is, we should fix the variables and the level of the approximation
{\em before} the calculations, we cannot do it on the fly.

\item A given, fixed state space and constitutive quantities still
does not warrant that the entropy production can be written in a
special quadratic form, as a sum of products of constitutive and
fixed functions of the state space. One should investigate and
clarify what are the conditions of that force-current structure.
We will see, that in classical treatments (classical irreversible
thermodynamics, classical extended irreversible thermodynamics)
the form of the entropy current function is an important condition
and we will investigate whether there are other choices in the
considered particular systems.
\end{itemize}

In this section we treat the above questions in a systematic treatment
We accept that the entropy is an extensive quantity, therefore there is
an entropy balance and we postulate that the Second Law ensures a
nonnegative entropy production. We will determine what are the state
variables and what are the constitutive quantities in a thermodynamic
theory. E.g. the entropy and the entropy current are always
constitutive, their evolution is determined by the dynamics of other
physical quantities. Our task is to get restrictions that a positive
entropy production and the other known restrictions (e.g. balance like
dynamics) results in the form of constitutive functions, hence in the
dynamics of the physical quantities. We apply the Liu theorem, given
shortly in a convenient form in Appendix A. First we treat classical
theories and we generalize them extending the constitutive spaces to
include nonlocal effects.

We suppose that the continuum is characterized by physical quantities
given by variables defined on the {\em basic state space} $Z$ and
therefore a local basic state of the continuum is given by ${\bf z} \in
Z$. The basic state space is called in a slightly different context as
the space of {\em wanted fields} \cite{MusAta01a}. It is convenient and
sufficiently general to assume that the basic state space is a
finite-dimensional vector space. The processes are space-time functions
on the basic state space. Therefore in a nonrelativistic treatment, for
a given observer a process is ${\bf z}_p: \mathbb{R}\times \mathbb{R}^3
\rightarrow Z$, $(t,{\bf x}) \mapsto {\bf z}_p(t,{\bf x})$. We are to
give the dynamic equations determining the processes, therefore we omit
the treatment of the structure of the process space, because it is
determined by the dynamic equations. The processes are supposed to be
continuously differentiable as many times as it is necessary. Moreover,
we are fully aware of the fact that the structure of space-time is more
refined also in the nonrelativistic case \cite{Mat84b,Mat85a}, but for
the sake of simplicity and easier understanding we restrict ourselves
to the level of treatment usual in the literature of the field. One
should be careful in applying our following considerations to
mechanical theories, because mechanics is strongly connected to the
structure of space-time \cite{BalVan02m}. To emphasize formally the
necessary care, the balances are written in a local frame, relatively
to an inertial observer. Without any further considerations they are
valid in a continuum at rest respectively to that observer.

We postulate that the entropy is an extensive quantity with non
negative production. This can be formulated in the form of a local
balance equation
\begin{equation}
\partial_t{s} + \nabla\cdot {\bf j}_s = \sigma_s \geq 0.
\label{BalEnt1}\end{equation}

$\partial_t$ denotes the partial derivative with respect to time. The
entropy density $s$ and the entropy current ${\bf j}_s$ are considered
as constitutive quantities, hence the nonnegative entropy production
$\sigma_s$ have to be a consequence of pure material properties. We
assume that the above inequality expresses a pure material property and
is fulfilled independently on the initial and boundary conditions of
the governing equations (see \cite{MusEhr96a}). We look for
constitutive relations with the help of Liu's theorem to ensure that
nonnegativity .

In addition of the entropy and the entropy current there can be
other constitutive quantities in the theory. The domain of the
constitutive functions is called {\em constitutive state space}.
The particular theories are determined basically by the choice of
the basic state space and the constitutive state space.

\subsection{Classical irreversible thermodynamics}

In classical irreversible thermodynamics the {\em basic state space}
$Z_{CIT}$ is spanned by the specific extensive variables. Therefore a
local basic state of the continuum is given as an element of this space
${\bf a} = (a_1,a_2,...,a_n) \in Z_{CIT}$, where $n$ is the number of
considered extensive quantities. The balance expressing the extensivity
of the variables is written as follows
\begin{equation}
\partial_t{\boldsymbol a} + \nabla\cdot {\bf j}_a = \sigma_a.
\label{BalExt1}\end{equation}

Here the current density of the extensives ${\bf j}_a=({\bf j}_1,{\bf
j}_2,...,{\bf j}_n)$ is a direct product vector of the conductive
current densities respectively and is a constitutive quantity. Every
current has a tensorial order one degree higher than the corresponding
specific extensive quantity. The dot between the nabla and the current
denotes a contraction for that extra tensorial order. The constitutive
space of classical irreversible thermodynamics is spanned by the
extensives and their gradients $C_{CIT} = Z_{CIT} \times Lin(Z_{CIT},
\mathbb{R}^3)$. A {\em constitutive state} of the material is an
element of this space $({\bf a}, \nabla{\bf a})\in C_{CIT}$. The so
called {\em process direction space} is spanned by the next space and
time derivatives of the basic state: $(\partial_t {\bf a},
\partial_t\nabla{\bf a}, \nabla^2{\bf a})$, where $\nabla^2$
denotes the second derivative with respect to the space variables. The
constitutive quantities that depend on the space-time through the
constitutive space are the entropy, the entropy current and the current
density of the extensives $(s, {\bf j}_s, {\bf j}_a)$. The subset of
the constitutive space where they are defined is the {\em constitutive
domain} $D\subset C_{CIT}$ of the classical irreversible thermodynamic
material.

After these preliminary nominations we are looking for restrictions of
the constitutive functions posed by the balance form dynamic equation
and the nonnegativity of the entropy` production. Considering the
previously given constitutive dependencies we can write (\ref{BalExt1})
and (\ref{BalEnt1}) as
\begin{eqnarray}
\partial_t {\bf a} &+&
    \partial_1 {\bf j}_a \cdot \nabla{\bf a} +
    \partial_2 {\bf j}_a : \nabla^2{\bf a} - \sigma_a = {\bf 0},
    \label{DerBal}\\
\partial_1s \partial_t {\bf a} &+&
    \partial_2s\partial_t\nabla{\bf a} +
    \partial_1 {\bf j}_s \cdot \nabla{\bf a} +
    \partial_2 {\bf j}_s : \nabla^2{\bf a} \geq 0.
    \label{DerBalEnt}
\end{eqnarray}

Here $\partial_1$ denotes the partial derivative of the
constitutive quantities with respect to {\bf a}, by the first
member of the constitutive space. $\partial_2$ is the derivative
with respect to the second member, by $\nabla {\bf a}$. Now we are
looking for constitutive functions that will ensure the validity
of the entropy inequality for all possible process directions
enabled by the balances.

ApplYING Liu's theorem to the equations above (see Appendix A) the
process direction space spanned by $(\partial_t {\bf a},
\partial_t\nabla{\bf a}, \nabla^2{\bf a})$ corresponds to $\mathbb{V}$.
The space into which the balances are mapping, that is essentially the
basic state space $Z_{CIT}$ divided by the one dimensional vector space
of vectorized time \cite{Mat84b}, corresponds to $\mathbb{V}'$. The
application of the theorem results in that there exist $\Gamma \in
Z_{CIT}^*$ and the related Liu-equations are
\begin{eqnarray}
\partial_1 s - \Gamma &=& {\bf 0}, \\
\partial_2 s &=& {\bf 0}, \\
(\partial_2{\bf j}_s - \Gamma \partial_2{\bf j}_a)^s &=& {\bf 0}.
\label{Liu-eqIrr3}\end{eqnarray}

The superscript $s$ denotes the symmetric part of the corresponding
second order tensor. Moreover, the dissipation inequality will be
$$
\partial_1{\bf j}_s \cdot\nabla {\bf a} - \Gamma \partial_1{\bf
j}_a\cdot\nabla{\bf a} + \Gamma\sigma_a \geq 0.
$$

From the first two Liu-equations one can see immediately that the
entropy is a function of the basic state {\bf a} and the Lagrange
multiplier can be identified with the intensive variable defined
by $Ds = \partial_1s = \Gamma$. A solution of (\ref{Liu-eqIrr3})
results in that
\begin{equation}
{\bf j}_s({\bf a},\nabla{\bf a}) = \Gamma({\bf a}) {\bf j}_a({\bf
a},\nabla{\bf a}) + {\bf j}_0({\bf a}),
\label{EC_IrrTerm}\end{equation}

\noindent where ${\bf j}_0$ is an arbitrary function. From these
consequences of the Liu equations one can get the dissipation
inequality as
\begin{equation}
{\bf j}_a\cdot\nabla\Gamma + \nabla\cdot{\bf j}_0 \geq 0.
\label{DI_IrrTerm}\end{equation}

If $\sigma_a = {\bf 0}$ and ${\bf j}_0={\bf 0}$ then the entropy
current and the entropy production reduces to the well known
classical form of irreversible thermodynamics
$$
{\bf j}_s = \Gamma {\bf j}_a
$$

\noindent and
$$
{\bf j}_a({\bf a},\nabla{\bf a}) \cdot\nabla\Gamma \geq 0.
$$

The general solution of the above inequality gives the classical
Onsagerian conductivity relations (see Appendix C), where one can
identify the current densities of extensives as thermodynamic currents
and the gradients of the intensives as thermodynamic forces of
classical irreversible thermodynamics.
$$
{\bf j}_a = {\bf L}\nabla\Gamma
$$

\noindent with positive semidefinite conductivity matrix {\bf L}. In
general the conductivity matrix can depend on the constitutive state
and is called as quasilinear. In case of constant conductivity one can
say that the theory is strictly linear \cite{Gya77a}.

It is interesting to observe that the validity of classical
force-current relations is connected to the condition that in the
entropy inequality $\sigma_a \Gamma$ should be nonnegative separately
to the other parts of the inequality. That condition is not true in
general and is used to separate dissipative and nondissipative parts of
the source term $\sigma_a$ (For example in a thermomechanics of
point-masses one can distinct between damping, conservative and exiting
momentum sources, force fields.) To avoid the details of that problem
we demanded that $\sigma_a={\bf 0}$. The systems corresponding to the
above special case are sometimes called as {\em pure dissipative} and
play a certain role in connection to variational principles of
irreversible thermodynamics \cite{Gya69a,Van96a}.

Furthermore let us observe that the general entropy current and the
source differs from the classical expressions in the term ${\bf j}_0$.
Several authors suggest an additive supplement to the classical entropy
current (e.g. the celebrated {\bf K} vector of M\"uller \cite{Mul68a}).
In the following in connection to more general theories we will see
that such term can have a physical significance which also cannot
excluded here (e.g. thermodynamic theories of chemical reactions).

We have seen that in our calculations the classical form  of the
entropy production (the first term in the inequality above) {\em and}
the classical form of the entropy current were consequences of the
Second Law, without referring to any 'local equilibrium' arguments.
Formulating the possible functional dependencies of the corresponding
quantities and the usual (sometimes) implicit postulates (balance
structure, existence of entropy and the entropy inequality), we
encountered a clear distinction between postulates and consequences and
the limits of validity of the local equilibrium approximation. For
example one can see, that the second half of the principle of local
equilibrium is completely unnecessary, the validity of the equilibrium
state functions and the classical form of the entropy current are both
consequences of structural requirements.

The choice of the constitutive space looks like a crucial point in the
above derivation. Remarkable, that an extension of the constitutive
space with higher order space derivatives does not alter the previous
scheme. A short calculation shows that in the final Onsagerian
conductivity equations the conductivity coefficients can be functions
of the full constitutive space, that is they can depend on all the
higher order space derivatives, but without further assumptions the
form of the entropy production does not change. What can make a
difference is the extension of the basic state space as we will see in
the next subsection.

\subsection{Classical extended irreversible thermodynamics}

Extended thermodynamics is a nonequilibrium thermodynamic theory, where
the currents of traditional extensive quantities are introduced as
internal variables. The theory is sometimes called {\it extended
irreversible thermodynamics} \cite{JouAta92b, JouAta99a}, {\it extended
rational thermodynamics} \cite{MulRug98b} or {\it wave approach to
thermodynamics} \cite{Ver97b} depending on the details of the approach
and the school of thermodynamicians. These details are not too
important in our treatment. One of the advantages of using currents as
independent variables is that in this case the final equations can be
compared directly with results of the kinetic theory, with the moment
series expansion of the Boltzmann equation. This comparison gives a
remarkable insight into the material characteristics of gases and
(partially) of fluids.

In the following we comply the procedure of the previous subsection in
a somewhat shortened form. In our treatment the difference is that the
basic state space of classical irreversible thermodynamics is
supplemented by the space of current density of the extensives, it is
spanned by the variables $({\bf a}, {\bf j}_a) \in Z_{ET}$. Therefore,
an appropriate constitutive space will be $C_{ET} = Z \times
Lin(Z,\mathbb{R}^3)\times Lin(Z,\mathbb{R}^3)$ and a constitutive state
of the extended irreversible material is $({\bf a}, {\bf j}_a,
\nabla{\bf a})\in C_{ET}$. In this way we supposed that there should be
an evolution equation for the current and we are to find a possible
form of this equation, according to thermodynamic requirements. The
conditions are the balances of extensives and the nonnegative entropy
production. Therefore the balance (\ref{BalExt1}) and the inequality
(\ref{BalEnt1}) are supplemented by the next general form of the
searched evolution equation
\begin{equation}
\partial_t{\bf j}_a + \mathcal{G} = {\bf 0}.
\label{EvJ}\end{equation}

The constitutive quantities defined on the constitutive space are $s,
{\bf j}_s$ and $\mathcal{G}$. Considering the constitutive
dependencies, the balances and the entropy inequality appear in the
form
\begin{gather*}
\partial_t{\boldsymbol a} + \nabla\cdot {\bf j}_a = \sigma_a, \\
\partial_t{\bf j}_a + \mathcal{G} = {\bf 0},\\
\partial_1 s \partial_t {\bf a} +
    \partial_2 s\cdot\partial_t {\bf j}_a +
    \partial_3 s\cdot\partial_t\nabla{\bf a} +
    \partial_1 {\bf j}_s \cdot \nabla{\bf a} +
    \partial_2 {\bf j}_s : \nabla{\bf j}_a +
    \partial_3 {\bf j}_s \cdot : \nabla^2{\bf a} \geq 0.
\end{gather*}

Here, according to our concise notation, the different dots are
denoting the contractions of the spacelike tensorial orders (not
involved in $Z_{ET}$). Therefore, because of Liu's theorem there exist
Lagrange multipliers $\Gamma_e$ and $\Gamma_n$ of the first two
equations so that Liu equations are
\begin{eqnarray*}
\partial_1 s - \Gamma_e &=& {\bf 0}, \\
\partial_2 s - \Gamma_n &=& {\bf 0}, \\
\partial_3 s &=& {\bf 0}, \\
\partial_2 {\bf j}_s - \Gamma_e {\bf I} &=& {\bf 0}, \\
(\partial_3 {\bf j}_s)^s &=& {\bf 0}.
\end{eqnarray*}
The dissipation inequality in our case is
\begin{equation}
 \partial_1 {\bf j}_s\cdot \nabla {\bf a} - \Gamma_n \mathcal{G}
 -\Gamma_e \sigma_a\geq 0.
\label{Dis-exteq}\end{equation}

A solution of Liu equations gives that $s$ and ${\bf j}_s$ do not
depend on the gradient of {\bf a} and the partial derivatives of the
entropy give the Lagrange multipliers, $Ds({\bf a}, {\bf
j}_a)=(\partial_1 s, \partial_2 s)= (\Gamma_e,\Gamma_n)$. Thus the Liu
equations give a condition for the entropy current as
\begin{equation}
\partial_2 {\bf j}_s({\bf a},{\bf j}_a) =
    \partial_1 s({\bf a},{\bf j}_a)  I,
\label{ExtLiuSimp}\end{equation}

\noindent where $I$ is the unit tensor of the basic state space. The
postulate of pure dissipativity ($\sigma_a = {\bf 0}$) also simplify
the dissipation inequality and we get that
\begin{equation}
\partial_1{\bf j}_s\cdot \nabla {\bf a} - \partial_2s \cdot
\mathcal{G} \geq 0.
\label{DI_ExtIT}\end{equation}

However, we cannot give a general solution of (\ref{ExtLiuSimp}) and
(\ref{DI_ExtIT}) without any further ado. Fortunately we can go
farther, because according to theories of extended thermodynamics one
can exploit some physical requirements writing our non-equilibrium
entropy function in the following form
\begin{equation}
s({\bf a},{\bf j}_a) = s_0({\bf a}) - \frac{1}{2} {\bf j}_a\cdot
{\bf m}({\bf a},{\bf j}_a) \cdot{\bf j}_a,
\label{WaveEntropy}\end{equation}

\noindent where the matrix {\bf m} is called the {\em matrix of the
thermodynamic inductivities}, $s_0({\bf a})= s({\bf a}, {\bf 0})$ is
the {\em equilibrium entropy}, that depends only on the extensives. It
is worth remarking that this form, with state dependent thermodynamic
inductivities is not an approximation if we accept the following
requirements:
\begin{itemize}
\item The current variables ${\bf j}_a$ are nonequilibrium, dynamic
variables (dynamic degrees of freedom) that identically vanish in
(local) thermodynamic equilibrium. There are extensives with such
currents (heat and diffusion currents) but there are also evident
counterexamples (pressure). The treatment of nonlocality connected to
non dynamic variables is beyond the scope of this paper.
\item The entropy is a concave function. According to that
requirement it is concave also in the nonequilibrium part of the
state space therefore the symmetric part of ${\bf m}$ is assumed
to be positive definite.
\item The derivative of the entropy function characterizes the
equilibrium.
\end{itemize}

With this premises the mean value theorem results in
(\ref{WaveEntropy}) as the most general form of the entropy (see
Appendix B). In case of general internal variables where the physical
meaning is fixed only after the thermodynamic restrictions are applied,
the Morse lemma can give a more restricted form of the entropy
function, with constant thermodynamic inductivities
\cite{Ver97b,Gil81b}. However, in extended thermodynamics the current
densities cannot be transformed arbitrarily, therefore a constant
inductivity matrix would be an approximation.

The form (\ref{WaveEntropy}) of non-equilibrium entropy was introduced
first by Machlup and Onsager for discrete systems using the time
derivatives of the original $\alpha$ quantities as dynamic variables
\cite{MacOns53a}. It was Gyarmati who suggested the above form for
field theories using the current densities as dynamic variables
\cite{Gya77a}. This kind of entropy function was originally a
characteristic constituent of the wave approach of thermodynamics, the
other theories realized the principle of local state through a
generalization of the Gibbs relation.

The previously introduced non-equilibrium intensives are the following
\begin{eqnarray}
\Gamma_e({\bf a},{\bf j}_a) &:=& \frac{\partial s}{\partial\bf a} =
\frac{\partial s_0}{\partial \bf a} - \frac{1}{2} {\bf j}_a \cdot
\frac{\partial {\bf m}}{\partial \bf a}
\cdot {\bf j}_a, \\
\Gamma_n({\bf a},{\bf j}_a) &:=&
\frac{\partial s}{\partial {\bf j}_a} =
 -\left( {\bf m}^s + \frac{1}{2} {\bf j}_a \cdot \frac{\partial {\bf
    m}}{\partial {\bf j}_a} \right) \cdot {\bf j}_a =:
-\widehat{\bf m}\cdot {\bf j}_a, \label{NEqInt}
\end{eqnarray}

\noindent where the superscript $ ^s$ denotes the symmetric part of the
corresponding quantity and $\widehat{\bf m}$ is a nonequilibrium
inductivity. $\widehat{\bf m}$ is symmetric if ${\bf m}$ is constant.

The form (\ref{WaveEntropy}) is still too general to solve
(\ref{ExtLiuSimp}) and (\ref{DI_ExtIT}) without further assumptions,
therefore we restrict ourselves to the usual approximation, when the
matrix of thermodynamic inductivities is constant:
$$
s({\bf a},{\bf j}_a) = s_0({\bf a}) - \frac{1}{2}{\bf j}_a \cdot
{\bf m}\cdot {\bf j}_a.
$$

In this case the solution of (\ref{ExtLiuSimp}) is similar to that we
have got in the previous subsection
$$
{\bf j}_s({\bf a}, {\bf j}_a) = \partial_1 s {\bf j}_a + {\bf
j}_0({\bf a}).
$$

The dissipation inequality becomes
$$
(\mathcal{G}\cdot {\bf m}^s + \nabla \partial_1 s)\cdot {\bf j}_a \geq
0.
$$

We have got a product of fixed functions of the basic state and
constitutive quantities to be determined. The general solution of this
kind of inequalities is treated shortly in Appendix C. Here we have got
constitutive expressions for $\mathcal{G}$ as Onsagerian conductivity
relations. Therefore, the form of the evolution equation of the current
densities is determined by
$$
 {\bf j}_a = {\bf L}(\mathcal{G}\cdot {\bf m}^s + \nabla \partial_1
s).
$$

Here {\bf L} is a positive semidefinite conductivity  matrix, depending
on the constitutive state. Substituting into (\ref{EvJ}) we get
$$
\partial_t {\bf j}_a +
    {\bf m}^{s-1} ( {\bf L}^{-1}{\bf j}_a -
    \nabla \partial_1 s)  = {\bf 0}.
$$

\subsection{Weakly nonlocal extended irreversible thermodynamics}
\label{NEITex}

Our next example is the first weakly nonlocal extension of
extended irreversible thermodynamics. The basic state space is the
same as in the last subsection but we will consider an extension
of the constitutive space. Here we need the first and second
derivatives of the current densities, too. Hence the constitutive
space of nonlocal extended thermodynamics is $C_{NET} = Z \times
Lin(Z,\mathbb{R}^3)\times Lin(Z,\mathbb{R}^3)\times
Lin(Z,\mathbb{R}^6)\times Lin(Z,\mathbb{R}^9)$ and the
constitutive functions will depend on the quantities $({\bf a},
{\bf j}_a, \nabla{\bf a}, \nabla{\bf j}_a,\nabla^2{\bf j}_a)\in
C_{NET}$. The balance of the extensives (\ref{BalEnt1}), the
evolution equation of the currents (\ref{EvJ}) and the entropy
inequality are to be solved.

In classical irreversible thermodynamics and extended irreversible
thermodynamics the choice of the state space made possible to prove the
local equilibrium and local state hypothesis: the entropy function did
not depend on the nonlocal part of the constitutive space, on the
gradient of the extensive quantities. In our case, the above extension
of the state space does not enable such kind of proof, therefore we
will postulate it. This is an assumption, we will call that the {\em
hypothesis of relocalizability}. However, instead of directly
prescribing that the entropy does not depend on the gradients we make a
seemingly weaker hypothesis requiring that the gradient of the current
evolution equation (\ref{EvJ}) does not appear among the constraints.
Furthermore, we treat pure dissipative systems with dynamic current
densities.

The constitutive quantities defined on the constitutive space are the
same as previously $s, {\bf j}_s$ and $\mathcal{G}$. Accordingly, the
balances and the entropy inequality appear in the form
\begin{gather*}
\partial_t{\boldsymbol a} + \nabla\cdot {\bf j}_a = {\bf 0}, \\
\partial_t{\bf j}_a + \mathcal{G} = {\bf 0},\\
\partial_1s \partial_t {\bf a} +
    \partial_2s\cdot\partial_t{\bf j}_a +
    \partial_3s\cdot\partial_t\nabla{\bf a} +
    \partial_4s:\partial_t\nabla{\bf j}_a +
    \partial_5s\cdot:\partial_t\nabla^2{\bf j}_a +\\
    \partial_1 {\bf j}_s \cdot \nabla{\bf a} +
    \partial_2 {\bf j}_s : \nabla{\bf j}_a +
    \partial_3 {\bf j}_s : \nabla^2{\bf a} +
    \partial_4 {\bf j}_s \cdot : \nabla^2{\bf j}_a +
    \partial_5 {\bf j}_s :: \nabla^3{\bf j}_a \geq 0.
\end{gather*}

Introducing the Lagrange multipliers $\Gamma_e$ and $\Gamma_n$ for the
first two equalities, the Liu equations are
\begin{eqnarray*}
\partial_1 s - \Gamma_e &=& 0,\\
\partial_2 s - \Gamma_n &=& 0,\\
\partial_i s &=& 0, \quad i=3,4\\
(\partial_5 s)^s &=& 0, \\
(\partial_i {\bf j}_s)^s &=& 0, \quad i=3,5.
\end{eqnarray*}

Here the superscript $s$ of $\partial_5 {\bf j}_s$ can be expressed
best with indexes that $(\partial_5)_{ijk} ({\bf j}_s)_l$ is symmetric
in $ijk$. Because of the larger constitutive space the dissipation
inequality is considerably longer
\begin{equation}
    \partial_1 {\bf j}_s \cdot \nabla {\bf a} +
    (\partial_2 {\bf j}_s - \Gamma_e {I}):\nabla {\bf j}_a +
    \partial_4 {\bf j}_s \cdot :\nabla^2 {\bf j}_a -
    \Gamma_n \mathcal{G} \geq 0.
\label{Dis-nexteq}\end{equation}

A solution of the Liu equations gives that the entropy is a function of
{\bf a} and ${\bf j}_a$ solely and does not depend on the gradients at
all (that is $s({\bf a}, {\bf j}_a)$) and ${\bf j}_s$ does not depend
on the gradient of {\bf a} and the second gradient of ${\bf j}_a$ (that
is ${\bf j}_s({\bf a}, {\bf j}_a,\nabla{\bf j}_a)$). The partial
derivatives of the entropy give the Lagrange multipliers respectively.
The dissipation inequality reduces to
$$
\nabla \cdot {\bf j}_s - \Gamma_e \nabla {\bf j}_a \ - \Gamma_n
\mathcal{G} \geq 0.
$$

We can see that the solution of the Liu equations does not give a
soluble form of the dissipation inequality. However, it is easy to
transform it to the traditional force-current form after some further,
not too restrictive premises. Let us turn our attention to the entropy
current. A natural physical assumption is that entropy, being connected
to other extensives, does not flow if the extensive quantities do not.
Here we should aply the same assumption as in case of classical
extended thermodynamics suppose the the current of the extensives is a
{\em dynamic variable} and vanishes in local thermodynamic equilibrium.
Therefore we can assume that the entropy current is zero if the
currents of the extensive quantities are identically zero ${\bf
j}_s({\bf a},{\bf 0}, {\bf 0})={\bf 0}$. Considering this condition and
using the mean value theorem, one can get the next functional form
\begin{equation}
{\bf j}_s({\bf a}, {\bf j}_a, \nabla{\bf j}_a) = {\bf B}({\bf a}, {\bf
j}_a, \nabla{\bf j}_a)\cdot {\bf j}_a,
\label{EntrCurr}\end{equation}
where the {\em current intensity factor} {\bf B}, can be supposed to be
continuously differentiable in a neighborhood of the local equilibrium
state. We should remark again that this form itself is not an
approximation, it is general with the above conditions. One can observe
that the tensorial order of the current intensity factor ${\bf B}$ is
one order higher than the corresponding current. As we mentioned above,
the classical form of the entropy current was investigated and
generalized by several authors. The above kind of miltiplicative
generalization is due to Ny\'\i{}ri \cite{Nyi91a1}.

Now the entropy production can be written as
\begin{equation}
({\bf B} - \Gamma_e I): \nabla{\bf j}_a + (\nabla\cdot {\bf B} +
\mathcal{G} \widehat{\bf m})\cdot {\bf j}_a \geq 0.
\label{EntrProdWNExt}\end{equation}

\noindent where $\widehat{\bf m}$ is the inductivity function defined
in (\ref{NEqInt}). The general solution of the inequality is
\begin{eqnarray}
\nabla\cdot {\bf B} - \partial_t{\bf j}_a \cdot\widehat{\bf m} &=&
    {\bf L}_{11}{\bf j}_a + {\bf L}_{12}\nabla {\bf j}_a,
    \label{O-1}\nonumber\\
{\bf B} - \Gamma_e {I} &=&
     {\bf L}_{21}{\bf j}_a + {\bf L}_{22} \nabla {\bf j}_a.
\label{Ons2}
\end{eqnarray}

We can eliminate the current intensity factor and get
\begin{eqnarray}
\widehat{\bf m}^* \cdot \partial_t{\bf j}_a -
    \nabla\cdot\left( {\bf L}_{21} {\bf j}_a +
    {\bf L}_{22}\nabla {\bf j}_a + \Gamma_e {I} \right)
   &=& - \left( {\bf L}_{11}{\bf j}_a +
    {\bf L}_{12}\nabla {\bf j}_a \right).
\label{genco4nst}\end{eqnarray}

The star denotes the transpose. One can see, that contrary to the
common postulate \cite{MulRug98b} in extended thermodynamics, this
evolution equation does not have a balance form in  general. It could
be transformed formally into a balance form with special source in case
of constant inductivities.

Let us remark that a smaller extension of the constitutive space with
only the gradient of the current densities would not have been enough
to get a gradient dependent entropy current.

\subsection{Weakly nonlocal internal variables}

In the previous subsections the physical meaning of the internal
variables was fixed, they were the current densities of the
thermodynamic extensives. However, it is a restriction, the usage and
applicability of internal variables is far more broader
\cite{MauMus94a1,MauMus94a2}. We can meet them under different names as
{\em order parameters, dynamic degrees of freedom, dynamic variables}
in different physical theories with a slightly different meaning. They
are applied characterizing materials with complex microstructure,
properties of gases near critical states, etc. The extent and method of
applying thermodynamic restrictions in internal variable theories can
be very different in different approaches. In thermodynamic theories
the application of internal variables is connected to the principle of
local state and therefore to discrete systems \cite{ColGur67a}. Their
dynamics is restricted or generated constitutively by the Second Law
and results in ordinary differential equations. In discrete systems,
i.e. in homogeneous bodies, the formalism is well known and frequently
applied, however sometimes the dynamic aspects are hidden (a good
summary of the dynamics of the simplest discrete thermodynamic system
without internal variables is given from several point of view in
\cite{OrlRoz84a1,OrlRoz84a2,Mat92a1} and with internal variables in
\cite{Van95a,MusDom96a}). On the other hand, as we mentioned in the
introduction, these internal variable theories can be considered as a
"relocalization" of some memory, inertial effects of the equilibrium
variables. The homogeneous dynamics (ordinary differential equations)
in continuous systems has some important disadvantages: the local state
seems to be too tight from several points of view. First of all some
physical phenomena cannot be modelled with them (e.g. thickness of
shear bands, structure of surfaces, ...). Moreover, the numerical
solution of coupled ordinary and partial differential equations depends
on the discretization, on the mesh.

The situation is entirely different in case of weakly nonlocal
equations. Here one can distinguish between essentially two independent
approaches. One of them is based on the fundamental {\em balance
equations} introducing gradient dependent terms directly into the
constitutive functions, which are generated using the local equilibrium
or the local state hypothesis (see e.g. \cite{Fal92a} and numerous
examples in rheology and mechanics in general
\cite{ZbiAif88a1,VarAif94a,Bed00a}).

The other approach is even more intuitive. There one assumes that the
local entropy function depends on the gradients of the corresponding
physical quantities (usually on a rather simple way) and a kind of
'nonlocal intensives' are generated by the functional derivatives of
the entropy function. This method is used to get the Ginzburg-Landau
equation for a so called 'non conservative' physical quantity (internal
variable), and the Cahn-Hilliard equation for a 'conservative',
extensive one (model A and B with the terminology of Hohenberg and
Halperin \cite{HohHal77a}). The general equation for the nonequilibrium
reversible-irreversible coupling (GENERIC) is based on a generalization
of this kind of approach \cite{GrmOtt97a,OttGrm97a}.

The relation of the two approaches is not clear at all. With the
traditional approach (based on local equilibrium and state) one cannot
get the most important nonlocal theories of contemporary physics (e.g.
Ginzburg-Landau) on the other hand the variational derivations are too
intuitive and they have nothing to do with the balances of the
fundamental physical quantities. Criticizing the mentioned derivations
of Ginzburg-Landau and Cahn-Hilliard equations, Gurtin emphasizes the
importance of the separation of balances from the constitutive
properties:

{\em "My view is that while derivations of the form ... are useful and
important, they should not be regarded as basic, rather as precursors
of more complete theories. While variational derivations often point
the way toward a correct statement of basic laws, to me such
derivations obscure the fundamental nature of balance laws in any
general framework that includes dissipation." \cite{Gur96a} }

Gurtin himself uses a method exploiting his "microforce balance
principle", which is a new principle again, and we can apply Gurtin's
further critical remarks used against the variational approach also to
his propositions:
\begin{itemize}
\item The derivations limit the manner how other terms (e.g. rate
terms) enter the equations.
\item The derivations are specific, it is not clear how they can  be
generalized for general thermodynamic processes including not only
mechanics, diffusion or the dynamics of the order parameters near
critical states.
\end{itemize}

In this section we will see how far can we go with relocalized
theories, with the methods of irreversible thermodynamics.  We
introduce nonlocal internal variables connected to the entropy current.
First we will see the weakly nonlocal irreversible thermodynamic theory
of pure dynamic internal variables without further constraints, after
that we investigate the case of a nonlocal extension of a thermodynamic
theory containing also extensive variables.

\subsubsection{Weakly nonlocal dynamic internal variable -
Ginzburg-Landau equation}\label{GLex}

Our task is to find weakly nonlocal evolution equation of an internal
variable that corresponds to the requirement of nonnegative entropy
production. The basic state space, $Z_I$, is spanned by the vector
$\boldsymbol{\xi}$, denoting an array of internal variables with
different tensorial character. We are to find an evolution equation of
the internal variable in the form:
\begin{equation}
\partial_t{\boldsymbol{\xi}} + \mathcal{F} = {\bf 0}.
\label{EvXi}\end{equation}

The constitutive space is spanned by $\boldsymbol{\xi}$ and its first
and second gradients: $(\boldsymbol{\xi}, \nabla\boldsymbol{\xi},$
$\nabla^2\boldsymbol{\xi})$. Therefore $C_{I} = Z_\xi \times
Lin(Z_\xi,\mathbb{R}^3)\times Lin(Z_\xi,\mathbb{R}^6)$ is the
constitutive space of the nonlocal dynamics of an arbitrary internal
variable. The constitutive quantities are the entropy, the entropy
current and the form of the evolution equation $(s,{\bf j}_s,
\mathcal{F})$. Therefore, the positive entropy production is
supplemented by (\ref{EvXi})  in the Liu procedure.
\begin{gather*}
\partial_t{\boldsymbol{\xi}} + \mathcal{F} = {\bf 0},\\
\partial_1s \partial_t \boldsymbol{\xi} +
    \partial_2s\cdot\partial_t\nabla\boldsymbol{\xi} +
    \partial_3s:\partial_t\nabla^2\boldsymbol{\xi} +
    \partial_1 {\bf j}_s \cdot \nabla\boldsymbol{\xi} +
    \partial_2 {\bf j}_s : \nabla^2\boldsymbol{\xi} +
    \partial_3 {\bf j}_s \cdot :\nabla^3\boldsymbol{\xi} \geq 0.
\end{gather*}

Here we applied the hypothesis of relocalizability, as in case of
extended thermodynamic systems. According to Liu's theorem there exist
a $\Gamma$, to be determined from the Liu equations, which can be
written in a particularly simple form
\begin{eqnarray*}
\partial_1 s - \Gamma &=& {\bf 0}, \\
\partial_2 s &=& {\bf 0}, \\
\partial_3 s &=& {\bf 0}, \\
\partial_3 {\bf j}_s &=& {\bf 0}.
\end{eqnarray*}

The dissipation inequality in our case is
\begin{equation}
\partial_1 {\bf j}_s \cdot\nabla \boldsymbol{\xi} +
\partial_2 {\bf j}_s :\nabla^2 \boldsymbol{\xi}- \Gamma \mathcal{F} \geq 0.
\label{Dis-GLeq}\end{equation}

The solution of Liu equations gives that $s$ depends only on the
internal variable $\boldsymbol{\xi}$, $\Gamma = Ds(\boldsymbol{\xi})$
is the derivative of the entropy and ${\bf j}_s$ does not depend on the
second gradient of $\boldsymbol{\xi}$. Unfortunately these
considerations do not simplify the entropy inequality at all, we should
look for additional conditions. As previously we assume that
$\boldsymbol{\xi}$ is a dynamic variables, Therefore there is no flow
of entropy and the derivative of the entropy is zero if the internal
variables are zero, that is ${\bf j}_s({\bf 0}, {\bf 0})= {\bf 0}$ and
$\Gamma({\bf 0})= {\bf 0}$. The first condition is quite natural. The
second condition requires that the zero derivatives characterize the
equilibrium of the material. Thus we can write the entropy current in
the following form
$$
{\bf j}_s(\boldsymbol{\xi},\nabla\boldsymbol{\xi}) = {\bf
A}(\boldsymbol{\xi}, \nabla\boldsymbol{\xi}) \Gamma(\boldsymbol{\xi}).
$$

This form of the entropy current is general with the assumptions above.
As previously, we assume, that {\bf A} is a continuously differentiable
function. In this case the entropy inequality turns out to have the
form
$$
(\nabla\cdot{\bf A} - \mathcal{F})\Gamma + {\bf A}\cdot \nabla\Gamma \geq
0.
$$
Let us remark, that in (\ref{Dis-GLeq}) there are two constitutive
quantities (the entropy current and $\mathcal{F}$) and three additive
terms. To simplify the inequality for constructing a force current
system there is no other choice that we have done here: we should unite
two of the terms with some reasonable physical assumption. Now we can
solve the resulted inequality and get an Onsagerian structure.
\begin{eqnarray}
\nabla\cdot {\bf A} - \mathcal{F} &=& {\bf L}_{11} \Gamma +{\bf L}_{12}
    \nabla\Gamma, \label{GLO-1}\\
{\bf A} &=& {\bf L}_{21} \Gamma +{\bf L}_{22}\nabla\Gamma.
\label{GLO-2}\end{eqnarray}

Here {\bf L} is the positive semidefinite constitutive conductivity
matrix. Eliminating {\bf A} from (\ref{GLO-1}) and (\ref{GLO-2}), we
get
\begin{equation}
\partial_t{\boldsymbol{\xi}} +
  \nabla\cdot({\bf L}_{21} \Gamma +{\bf L}_{22}\nabla\Gamma) =
  {\bf L}_{11} \Gamma +{\bf L}_{12}\nabla\Gamma.
\label{genGL}\end{equation}

This is a balance equation with a special current and source term. To
recognize the structure of the equation let us consider a single scalar
internal variable $\xi$ and isotropic material. In this case there is
no cross effect between the thermodynamic forces of different tensorial
order ${L}_{12}={0}$ and ${\bf L}_{21}={\bf 0}$, and the diagonal
elements of the conductivity tensor ${\bf L}$ are scalars ${L}_{11}=
l_1$ and ${\bf L}_{22}= l_2 {I}$ (according to Curie principle or more
properly the representation theorems of isotropic tensors). Therefore
equation (\ref{genGL}) simplifies to
\begin{equation}
\partial_t\xi = l_1 \Gamma - \nabla\cdot(l_2 \nabla \Gamma_\xi).
\label{Ginzburg-Landau}\end{equation}

Here we have got an equation that is similar to the well known
Ginzburg-Landau equation. However, there are some differences.
\begin{itemize}
\item In the second term of the right hand side under the space
derivatives is $\Gamma$ instead of $\xi$, contrary to the
Ginzburg-Landau equation. The consequence of this difference is that
the equation has the traditional homogeneous thermodynamic equilibrium
solutions characterized by the condition $\Gamma=0$.
\item The sign of the material coefficients is determined by the Second
Law, not by direct stability considerations.
\item It is straightforward to extend the derivation to nonlinear and
anisotropic cases. However, the nonlinearities and anisotropies show a
different structure than in the original Ginzburg-Landau equation.
\end{itemize}

In the following (\ref{Ginzburg-Landau}) will be called {\em
thermodynamic Ginzburg-Landau equation} \cite{Van02a1}. This equation
was derived  by Verh\'as under some slightly different assumptions, as
a governing equation for the transport of dynamic degrees of freedom
\cite{Ver83a,Ver96a,Ver97b}.

\begin{rem} In all of the previous considerations the only
mathematical step that seems to be restrictive from a physical point of
view is the (continuous) differentiability of the current intensity
factor. That condition can be weakened, but it will exclude the
appearance of terms like ${\bf j}_0$ in case of classical and extended
thermodynamics. The question is not mathematical and academic at all.
We should not forget that the question of differentiability plays an
important role in constitutive modelling and in thermodynamics in
general. For example some phase boundaries are classified according to
the differentiability of equilibrium state functions. Moreover,
recently an experimental indication of phase boundaries was found also
in the non-equilibrium state space, connected to internal variables.
That concept makes possible the unification and the elaboration of a
common thermodynamic frame for theories of failure and for theories of
elasticity in mechanical continuum (e.g. in case of damaged brittle
materials \cite{Van01a1,VanVas01p}).
\end{rem}

\subsubsection{Weakly nonlocal dynamic internal variable and extensive
variable - Cahn-Hilliard equation}\label{CHex}

In our last example there is an extensive thermodynamic variable and
also an internal one that is not the current density of the extensive
as in extended irreversible thermodynamics. Therefore the basic state
space of the physical quantities consists of the extensive quantity
${\bf a}$ with a balance like evolution equation (\ref{BalExt1}) and a
dynamic internal variable $\boldsymbol{\xi}$ with an evolution equation
(\ref{EvXi}), whose form is not specified. We are looking for nonlocal
extension of the internal variable, therefore the constitutive space
consists of the constitutive spaces of classical irreversible
thermodynamics and the previous pure nonlocal internal variable theory
together $C_{IE} = Z_a \times  Z_\xi \times Lin(Z_a,\mathbb{R}^3)
\times Lin(Z_\xi,\mathbb{R}^3)\times Lin(Z_\xi,\mathbb{R}^6)$. Our
constitutive quantities are the entropy, the entropy current, the
current of the extensives and the evolution equation of the internal
variable $(s, {\bf j}_s, {\bf j}_a ,\mathcal{F})$. All these quantities
are functions defined on the constitutive space spanned by $({\bf a},
\boldsymbol{\xi}, \nabla{\bf a}, \nabla\boldsymbol{\xi},
\nabla^2\boldsymbol{\xi})$. We are looking for the restrictions on the
constitutive quantities imposed by the nonnegative entropy production
(\ref{BalEnt1}) and the balance (\ref{BalExt1}) with zero source term
(pure dissipativity) and the condition of relocalizability. Considering
the later we do not impose the space derivative of the evolution
equation of the internal variables as an additional constraint. Because
of the constitutive dependencies, the constraints and the entropy
inequality can be written in the following form
\begin{gather*}
{\partial_t\boldsymbol{\bf a}} +
    \partial_1 {\bf j}_a \cdot \nabla{\bf a} +
    \partial_2 {\bf j}_a \cdot \nabla\boldsymbol{\xi} +
    \partial_3 {\bf j}_a : \nabla^2{\bf a} +
    \partial_4 {\bf j}_a : \nabla^2\boldsymbol{\xi} +
    \partial_5 {\bf j}_a \cdot : \nabla^3\boldsymbol{\xi}
    = {\bf 0},\\
\partial_t{\boldsymbol{\xi}} + \mathcal{F} = {\bf 0},\\
\partial_1s \partial_t {\bf a}  +
    \partial_2s \partial_t\boldsymbol{\xi} +
    \partial_3s \cdot\partial_t \nabla {\bf a} +
    \partial_4s \cdot\partial_t\nabla\boldsymbol{\xi} +
    \partial_5s :\partial_t\nabla^2\boldsymbol{\xi} + \\
    \partial_1 {\bf j}_s \cdot \nabla{\bf a} +
    \partial_2 {\bf j}_s \cdot \nabla\boldsymbol{\xi} +
    \partial_3 {\bf j}_s  : \nabla^2{\bf a} +
    \partial_4 {\bf j}_s  : \nabla^2\boldsymbol{\xi} +
    \partial_5 {\bf j}_s \cdot :\nabla^3\boldsymbol{\xi}  \geq 0.
\end{gather*}

According to Liu's theorem there exist functions $\Gamma_e$ and
$\Gamma_\xi$ to be determined from Liu equations written as
\begin{eqnarray*}
\partial_1 s - \Gamma_e &=& {\bf 0}, \\
\partial_2 s - \Gamma_\xi &=& {\bf 0}, \\
\partial_i s &=& {\bf 0}, \quad i=3,4, \\
(\partial_5 s)^s &=& 0, \\
(\partial_3 {\bf j}_s - \Gamma_e \partial_3{\bf j}_a)^s &=& {\bf 0}, \\
(\partial_5 {\bf j}_s - \Gamma_e \partial_5{\bf j}_a)^s &=& {\bf 0}.
\end{eqnarray*}

Here the superscript $s$ of the last equation can be expressed best
with indexes that $(\partial_5)_{ijk} ({\bf j})_l$ is symmetric in
$ijk$. A solution of the Liu equations results in that the entropy
depends only on the extensive and on the internal variable. The
derivative of the entropy gives the Lagrange multipliers  $Ds({\bf a},
\boldsymbol{\xi}) = (\partial_a s, \partial_\xi s) = (\Gamma_e,
\Gamma_\xi)$. A consequence of the last two equations is
\begin{equation}
{\bf j}_s = \Gamma_e{\bf j}_a + {\bf j}_0({\bf a}, \boldsymbol{\xi},
\nabla\boldsymbol{\xi}).
\label{CHEntCurr}\end{equation}

Here ${\bf j}_0$ is an arbitrary function of the variables denoted
above. In the first term we can recognize the classical entropy
current. This solution of the Liu equations simplifies the dissipation
inequality to
\begin{equation}
\nabla \Gamma_e \cdot {\bf j}_a + \nabla\cdot{\bf j}_0 -\Gamma_\xi
\mathcal{F} \geq 0. \label{Dis-CHeq}\end{equation}

This form is simple but unfortunately we cannot solve it withoutany
forther ado. We require that in case of zero internal variable the
entropy flow reduces to the classical form. The dynamic property of the
internal variable gives the additive term of the entropy current in the
following form
$$
{\bf j}_0({\bf a}, \boldsymbol{\xi}, \nabla\boldsymbol{\xi}) = {\bf
A}({\bf a}, \boldsymbol{\xi}, \nabla\boldsymbol{\xi})\Gamma_\xi.
$$
Now the dissipation inequality transforms to
$$
{\bf j}_a\cdot \nabla\Gamma_e +
    {\bf A}\cdot\nabla\Gamma_\xi +
    (\nabla\cdot{\bf A} - \mathcal{F})\Gamma_\xi  \geq 0.
$$

As in the previous sections we reckon the entropy as a primary
quantity. Therefore we arrived to a force-current system, to a sum of
products of undetermined and given functions of the constitutive state.
Therefore the general solution the inequality results in the Onsagerian
conductivity equations in a quasilinear form, with a positive
semidefinite conductivity matrix {\bf L}, depending on the constitutive
state
\begin{eqnarray}
{\bf j}_a
&=&
    {\bf L}_{11} \nabla \Gamma +
    {\bf L}_{12} \nabla \Gamma_{\xi} +
    {\bf L}_{13} \Gamma_\xi,    \label{CHOq-1}\\
{\bf A} &=&
    {\bf L}_{21} \nabla\Gamma +
    {\bf L}_{22} \nabla\Gamma_{\xi} +
    {\bf L}_{23} \Gamma_\xi,    \label{CHOq-2}\\
\nabla\cdot{\bf A} - \mathcal{F} &=&
    {\bf L}_{11} \nabla \Gamma +
    {\bf L}_{12} \nabla \Gamma_{\xi} +
    {\bf L}_3 \Gamma_\xi.    \label{CHOq-3}
\end{eqnarray}

Substituting (\ref{CHOq-1}) into (\ref{BalExt1}) and (\ref{CHOq-2})
into (\ref{CHOq-3}) we arrive the following system of transport like
equations with a definite balance form
\begin{eqnarray}
\partial_t{\bf a} +
    \nabla\cdot\left({\bf L}_{11} \nabla \Gamma +
        {\bf L}_{12} \nabla \Gamma_{\xi} +
        {\bf L}_{13} \Gamma_\xi \right) &=& {\bf 0}, \label{TranEx}\\
\partial_t\boldsymbol{\xi} +
    \nabla\cdot\left( {\bf L}_{21} \nabla\Gamma +
        {\bf L}_{22} \nabla\Gamma_{\xi} +
        {\bf L}_{23} \Gamma_\xi\right)&=&
        {\bf L}_{11} \nabla \Gamma +
        {\bf L}_{12} \nabla \Gamma_{\xi} +
        {\bf L}_3 \Gamma_\xi.    \label{TranInt}
\end{eqnarray}

On the other hand this choice of the basic state space indicates that
the above set of evolution equations can be considered as
generalization of the Cahn-Hilliard equation. To see the connection one
must eliminate the internal variable. Let us consider a very special
example, where our basic state space is spanned by a single scalar
extensive $a$ and a single scalar internal variable $\xi$. Furthermore,
let be the material isotropic and all the material coefficients are
constant (strictly linear approximation). Finally, let us write the
entropy in the canonical form (\ref{WaveEntropy}), but with constant
inductivities $m$. Now there are two coupled terms in the linear laws,
because the current of the extensives ${\bf j}_a$ and the current
multiplier {\bf A} have same tensorial order
\begin{eqnarray}
{\bf j}_a &=& l_{11} \nabla \Gamma - l_{12} m \nabla {\xi},
    \label{CHO-1}\\
{\bf A} &=& l_{21} \nabla\Gamma - l_{22} m \nabla {\xi} ,
    \label{CHO-2}\\
\nabla\cdot{\bf A} - \mathcal{F} &=& - l_3 m{\xi}.
    \label{CHO-3}
\end{eqnarray}

Here the matrix of the conductivity coefficients is positive definite
to ensure nonnegative entropy production ($l_3 \geq 0$, $l_{11} \geq
0$, $l_{22} \geq 0$ and $l_{11} l_{22} - l_{12} l_{21} \geq 0$). We can
eliminate $\nabla{\xi}$ from the first two equations to get an
expression of the current multiplier {\bf A} as:
$$
{\bf A} = l_{22} l_{12}^{-1} {\bf j}_a + (l_{21} - l_{22} l_{12}^{-1}
l_{11})\nabla \Gamma.
$$

We can put this formula into (\ref{CHO-3}) and so eliminating ${\xi}$
from (\ref{CHO-1}) we get
\begin{equation}
{\bf j}_a = l_{11}\nabla (\Gamma + \alpha \Delta\Gamma) + \beta
\nabla\nabla\cdot {\bf j}_a + l_{12}l_3^{-1} \partial_t{\xi},
\label{jeq}\end{equation}

\noindent where $ \alpha = l_{11}^{-1}l_3^{-1}(l_{12}l_{21}-l_{22}
l_{11})$ and $\beta =l_3^{-1} l_{22}$. we can neglect the last term in
the above expression if $l_{12}=0$ or the change of $\xi$ is very slow.
So we eliminate the internal variable and got (\ref{BalExt1}) and
(\ref{jeq}) as a system of equations to be solved. After some
straightforward manipulation we finally get
\begin{equation}
\partial_t{ a}+ l_{11}\Delta(\Gamma - \alpha \Delta \Gamma) - \beta \Delta
\partial_t{ a}= 0. \label{TCH}\end{equation}

In our case, when $a$ represents a single scalar physical quantity
(e.g. mass density) then $l_{11}$, $\alpha$ and $\beta$ are positive
numbers. Equation (\ref{TCH}) is similar to the Cahn-Hilliard equation,
but not the same, therefore it will be called {\em thermodynamic
Cahn-Hilliard equation}. It is worth to give a short comparison.
\begin{itemize}
\item The last extra term does not appear in the traditional equation.
Gurtin derived a similar additional term with the help of microforce
balance (a strong additional assumption) in the special case when the
variable $a$ is the mass density \cite{Gur96a}. As regards some
generalizations and the properties of the solution see
\cite{Mir00a,BonMir01a}. This term is similar that one could get in
case of the Guyer-Krumhansl equation and there are several experimental
indications for this kind of generalization with the Cahn-Hilliard
equation, too \cite{Aif80a1}. Supposing that $l_{22}=0$, this
additional term vanishes. Let us recognize that this rate dependent
term appears in the linear Onsagerian conductivity equations at the
same thermodynamic approximation level as all the others.
\item Instead of the extensive ${\bf a}$, the intensive variable
$\Gamma$ is under the second Laplace operator, contrary to the original
Cahn-Hilliard equation. The equilibrium solutions of the thermodynamic
Cahn-Hilliard equation are homogeneous.
\item Every material parameter in (\ref{TCH}) is positive, as a
consequence of the Second Law, a nonnegative entropy production. In the
original equation the signs of the coefficients are fixed according to
stability considerations seemingly independently of the entropy
balance.
\item The nonlinearities and anisotropies show a different structure
than in the original equation.
\item In most of the previous derivations of the Cahn-Hilliard equation,
a "generalized chemical potential" appeared by analogy, and was
inserted without any further ado into the mass balance. Here in our
derivation we have got a real generalization of the diffusion current
in the same sense and with the same method as in the original
irreversible thermodynamic approach.
\item Treating the derivation of the thermodynamic Cahn-Hilliard equation
a scalar extensive and a scalar internal variable was introduced. As
{\bf a} can denote the Cartesian product of several extensives, the
coupling in conductivity equations (\ref{CHO-1})-(\ref{CHO-3}) can
become more involved, even in the simplest isotropic case. On the other
hand, the introduction of an internal variable with the tensorial order
of the current (one order higher than the extensive) results in the
very same structure of the final Cahn-Hilliard type equation.
\end{itemize}

\section{Heuristic weakly nonlocal irreversible thermodynamics}

In the previous section weakly nonlocal thermodynamics was treated in a
more or less mathematical level. Our point of view, the leading idea
behind the previous section was to give the background and the
conditions of applicability of the classical irreversible thermodynamic
approach far beyond the validity of local equilibrium. We have seen,
that the essence of the classical method of irreversible
thermodynamics, the construction of proper force-current systems can be
applied not only to systems with inertia, but for essentially nonlocal
systems, too. Moreover, the local character of the theory can be
preserved introducing special internal variables, current multipliers.

The general basic conditions to create solvable force-current systems
were the following:

\begin{enumerate}
\item {\em Entropy} - there exist an extensive entropy with
nonnegative production. Entropy and entropy current are constitutive
functions. The entropy is the primary constitutive function of any
thermodynamic theory.
\item {\em Pure dissipative systems} - the source term is zero
in the balances of considered extensives.

\item {\em Relocalizability} - the entropy, as fundamental potential
function expressing the properties of ordinary (equilibrium
thermodynamic) systems does not depend on gradients of the basic
variables.

In weakly nonlocal theories this is a postulate, in case of more
restricted constitutive spaces is a consequence of the thermodynamic
requirements as we have seen in the previous sections.

\item {\em Dynamic internal variables} - are those that vanish in
(local) equilibrium. This assumption results in two consequences for
sufficiently smooth constitutive functions:
\begin{itemize}
\item If the entropy is a concave function of the internal variables
(in the nonequilibrium part of the state space) then we arrive at the
Gyarmati form of nonequilibrium entropy (\ref{WaveEntropy}).
\item The Ny\'\i{}ri form generalized entropy current (\ref{EntrCurr}).
\end{itemize}\end{enumerate}

Here, in this section, based on the above properties, we treat some
further weakly nonlocal relocalizable systems in a more heuristic way.
This heuristic point of view shows, how one can get solvable (Liu)
equations and solvable entropy inequality in the more exact and formal
mathematical procedure. In the first subsection we investigate the
parade ground of thermodynamic theories and develop weakly nonlocal
heat conduction. After that we show that a current multiplier of
current multipliers can have an importance and derive Gurtin's term in
Ginzburg-Landau equation. Finally we recall that current multipliers
are not unique, there are several possibilities to transform the
dissipation inequality into a solvable form. Here we investigate
extended thermodynamics as a special internal variable theory and
investigate the conditions to get a balance like evolution equation for
the internal variables.

\subsection{Weakly nonlocal heat conduction - Guyer-Krumhansl equation
\cite{GuyKru66a1,GuyKru66a2}}

The different special theories of thermodynamic origin as diffusion,
mechanical, electromagnetic interactions, etc. all can show phenomena
that can be explained by weakly nonlocal extensions of the classical
equations. Let us investigate more closely the most frequently treated
basic example of all thermodynamic investigations, the phenomena of
heat conduction. Now the single extensive quantity is the internal
energy, therefore ${\bf a} = (a_1) \equiv u$. Instead of citing the
previously derived formulas we will apply a simplified, more heuristic
procedure. The example demonstrates the role of the continuization
hypotheses of local equilibrium, local state and relocalizability
giving the different heat conduction theories respectively. We will see
that the Guyer-Krumhansl equation is the thermodynamically consistent
first weakly nonlocal extension of the Cattaneo-Vernotte equation and
the structure of the equations is independent of the particular
material properties.

The balance of internal energy is the starting point. Let us denote the
heat current density by ${\bf j}_a = ({\bf j}_u \equiv) {\bf q}$ and we
assume that the system is purely dissipative.
\begin{equation}
\partial_t{u} + \nabla\cdot {\bf q} = 0.
\label{BalExt1Heat}\end{equation}

For the case of simplicity, we consider isotropic material and constant
thermal inductivity ${\bf m}$. In this case  ${\bf m}= m {I}$,
therefore the entropy function (\ref{WaveEntropy}) can be written as
\begin{equation}
s(u,{\bf q}) = s_0(u) - \frac{1}{2} m {\bf q}^2,
\label{HeatWaveEntropy}\end{equation}

\noindent where $m>0$. The Ny\'\i{}ri form of the entropy current
is
\begin{equation}
{\bf j}_s = {\bf B}_u \cdot {\bf q}.
\label{HeatEntrCurr}\end{equation}

Now the entropy production (\ref{EntrProdWNExt}) can be calculated by
composite derivations
\begin{gather}
\partial_t s + \nabla\cdot{\bf j}_s = \nonumber\\
\partial_us \partial_t u + \partial_qs \partial_t {\bf q} +
\nabla\cdot({\bf B}_u \cdot {\bf q}) = \nonumber\\
({\bf B}_u - \frac{1}{T} {I}):\nabla {\bf q} +
(\nabla\cdot {\bf B}_u - m \partial_t {\bf q} )\cdot {\bf q} \geq 0.
\label{HeatEntrProd}\end{gather}

We can recognize (\ref{EntrProdWNExt}) in our special case. According
to Curie principle (representation theorems of isotropic tensors, e.g.
\cite{PipRiv59a,Smi64a}) there is no cross coupling between the
different thermodynamic interactions. Moreover in (\ref{Ons2}) ${\bf
L}_{11}= l {I}$, where $l$ is a scalar and ${\bf L}_{22}$ can be
written with indices as
$$
({L}_{22})_{ijkl} = l_1\delta_{ik}\delta_{jl} +
l_2\delta_{il}\delta_{jk} + l_3 \delta_{ij}\delta_{kl}.
$$

Therefore the linear approximation of Onsager (\ref{Ons2}) reduces to
\begin{eqnarray}
{\bf q} &=& l (\nabla\cdot {\bf B}_u -  m \partial_t{\bf q})\nonumber\\
{\bf B}_u - \frac{1}{T} {\bf I} &=& l_1 (\nabla  {\bf q}) + l_2(\nabla
{\bf q})^* + l_3 \nabla \cdot{\bf q}{\bf I}.
\label{HeatO-2}\end{eqnarray}

\noindent where $*$ denotes the transpose. Eliminating ${\bf B}_u$ we
get
\begin{equation}
l m \partial_t{\bf q} + {\bf q} =
    l \nabla \frac{1}{T} + l (l_1 \Delta {\bf q} +
    (l_2+l_3)\nabla^2\cdot{\bf q}).
\label{IsoO-1}\end{equation}

This is exactly the Guyer-Krumhansl equation for the heat current,
introduced to describe the thermal properties of some crystals at low
temperatures \cite{GuyKru66a1}. Originally it was derived using kinetic
physics and later two-fluid hydrodynamics \cite{Enz74a}.  At the last
decades there were several attempts to get the equation from pure
non-equilibrium thermodynamics
\cite{Fek81a,Bha82p,LebDau90a,Net93a,LebGre96a,LebAta97a,LebAta98a}.
However, all of these derivations contain several ad-hoc assumptions,
whose generalization for more difficult situations can be very
ponderous if not impossible (e.g. two-fluid hydrodynamics is an
internal variable theory from a thermodynamic point of view, but with a
particular interpretation of the internal variables). One of these
assumptions is a balance like dynamics for the heat current {\bf q}
that is a result of the general structure in our case. If we do not
eliminate ${\bf B}_u$ from the material equations, then we can get a
dynamic equation of the heat current that has a special balance form
(more properly it will be an equation of Ginzburg-Landau type)
indicating that the current intensity factor can be considered as a
current density of the heat current {\bf q} \cite{Ver83a}. We analyze
further that property later.

In the special case of $l_1=l_2=l_3=0$, we get the Cattaneo-Vernotte
equation for the heat current. The conductivity equations indicate
clearly that the local state hypothesis is applied, the entropy current
has its traditional form ${\bf j}_s = {\bf j}_q/T$, but the entropy
itself depends on the dynamic nonequilibrium variable {\bf q}. The
Cattaneo-Vernotte equation is a wave like equation for the heat
conduction phenomena. It was Gyarmati who first derived it in the frame
of irreversible thermodynamics with the help of entropy function
(\ref{HeatWaveEntropy}) \cite{Gya77a} . Further, we get the Fourier
heat conduction equation when $m=0$, in local equilibrium, when the
entropy depends only on internal energy $u$.

In case of constant material coefficients ($m, l, l_1, ...$) it is easy
to derive a telegraph type equation directly for the temperature $T$.
We should introduce the specific heat $u=cT$ and  substitute
(\ref{BalExt1Heat}) and its time derivative into the divergence of
(\ref{IsoO-1})
\begin{equation}
l m c\partial_{tt}{T} + c\partial_t{T} + l \Delta \frac{1}{T} +
    l c(l_1+l_2+l_3) \Delta\partial_t{T} =0.
\label{GuKru}\end{equation}

This reduction is remarkable and discussed from several points of view.
First of all in several practical problems the material parameters
cannot be considered as constant quantities (see e.g.
\cite{ValAta97a}). On the other hand the reduction of a vector and a
scalar equation (\ref{BalExt1Heat}) and (\ref{IsoO-1}) into a single
scalar one results in a loss of information. The system
(\ref{BalExt1Heat}) and (\ref{IsoO-1}) is equivalent to (\ref{GuKru})
if the heat current field is rotation free.

\subsection{Nonlocal internal variables}

\subsubsection{Generalized thermodynamic Ginzburg-Landau equation}

In this subsection we develop the results of subsection \ref{GLex} and
show how can an additional rate term appear in the Ginzburg-Landau
equation.

As previously our task is to find an evolution equation of an internal
variable that corresponds to the requirement of nonnegative entropy
production. The basic state space is the vector (moreover Banach) space
of the variable $\boldsymbol{\xi}$ and we do not know a dynamic
equation for the processes. We suppose a relocalizable theory,
therefore the entropy function depends only on the internal variable.
We will use the notation $\Gamma_\xi := Ds(\boldsymbol{\xi})$. If
$\boldsymbol{\xi}=0$ then $\Gamma_\xi({\bf 0})={\bf 0}$, because
$\boldsymbol{\xi}$ is a dynamic variable. As regards the entropy
current we apply the previous physical assumptions also in this case:
if $\boldsymbol{\xi}$ was zero then there is no entropy flow. Moreover,
in the light of this assumption and according to the mean value
theorem, the entropy current can be written as a linear function of the
derivative of the entropy, as in classical irreversible thermodynamics
${\bf j}_s = {\bf A} \Gamma_\xi$. Here we introduced a nonlocal
internal variable, a current multiplier {\bf A}. We can go farther, and
introduce the current multiplier into the basic state space. For the
sake of simplicity, we suppose that the entropy function does not
depend on {\bf A}, that is we are not interested in the associated
memory effects, we are investigating only the nonlocal extension. The
form of the entropy current is similar to that of the Cahn-Hilliard
equation:
$$
 {\bf j}_s = {\bf A} \Gamma_\xi + {\bf B}\cdot {\bf A}.
$$

Here {\bf B} is a current multiplier of the first current multiplier
{\bf A}. This form expresses the physical assumption, that there is no
entropy flow when the internal variables $\boldsymbol{\xi}$ and {\bf A}
are zero. The tensorial order of {\bf A} is one order higher than the
tensorial order of $\boldsymbol{\xi}$ and the tensorial order of {\bf
B} is one order higher than the tensorial order of {\bf A} as it should
be considering the most general linear connection. Therefore the
entropy production follows as
\begin{equation}
\partial_t{s}(\boldsymbol{\xi}) + \nabla\cdot {\bf j}_s =
 (\partial_t{\boldsymbol{\xi}} +
 \nabla \cdot ({\bf A})\Gamma_\xi +
 {\bf A}\cdot (\nabla \Gamma_\xi +
 \nabla\cdot {\bf B}) +
 {\bf B}:\nabla {\bf A}  \geq 0.
\label{GLEnt2_ineq}\end{equation}

It is straightforward to put down the Onsagerian conductivity
equations, but in general one cannot simplify them. Therefore, we will
treat here only the simplest situation, when $\boldsymbol{\xi} = \xi$
is scalar, the material is isotropic and the approximation is strictly
linear (the conductivity coefficients are constants). Now the
conductivity equations are reduced to the following form
\begin{eqnarray}
\partial_t{{\xi}} + \nabla\cdot {\bf A} &=& l_1 \Gamma_\xi
\label{gGLO-1}\\
{\bf A} &=& l_2 (\nabla \Gamma_\xi + \nabla \cdot {\bf B})
\label{gGLO-2}\\
{\bf B} &=& l_3^1 \nabla {\bf A} + l_3^2 (\nabla {\bf A})^* +
l_3^3 \nabla\cdot {\bf A},
 \label{gGLO-3}\end{eqnarray}

\noindent where $l_1, l_2, l_3^1, l_3^2, l_3^3$ are positive,
scalar, constant coefficients. Now a simple calculation
eliminates {\bf A} and {\bf B} from the above equations:
\begin{equation}
\partial_t{{\xi}} = l_1 \Gamma_\xi -
    l_2\Delta(1+l_1 l_3)\Gamma_\xi +
    l_3 \Delta\partial_t{{\xi}}.
\end{equation}

Here $l_3 = l_3^1 + l_3^2 + l_3^3$. The last term, that is additional
to (\ref{Ginzburg-Landau}) corresponds to the generalized
Ginzburg-Landau equation of Gurtin. One can get back the thermodynamic
Ginzburg-Landau equation (\ref{Ginzburg-Landau}) in the special case
when $l_3^1=l_3^2=l_3^3=0$, that is ${\bf B}\equiv {\bf 0}$. The
positivity (positive definiteness in a more general situation) of the
material coefficients is ensured by the Second Law. However, we should
observe, that this generalization has not changed the characteristic
thermodynamic term which differs from the original Ginzburg-Landau
form: $\Gamma_\xi$ stands under the Laplace operator instead of
${\xi}$.

We can continue the introduction of new nonlocal internal variables,
putting {\bf B} into the basic state space. In this case ${\bf B}$
becomes an internal variable and we can introduce a corresponding
current multiplier. Continuing this procedure, we can develop a whole
phenomenological hierarchy of weakly nonlocal transport equations of
higher and higher orders. The further research in this direction has a
special importance for the kinetic theories. The outlined
phenomenological hierarchy of nonlocal equations can suggest an
approach similar to the momentum series expansion \cite{Lib90b,Net93a}.
On the other hand it is straightforward to extend the above treatment
(with Liu procedure or without it) considering memory effects, too.

\subsubsection{Are the dynamic equations of internal variables balances
or not?}

In section \ref{NEITex} we have seen that in extended irreversible
thermodynamics, when the currents of the extensives are the internal
variables, the final constitutive-evolution equation of the currents
did not have a balance form in general. Later, in section \ref{GLex}
for a pure internal variable the derived evolution equation
(\ref{TranInt}) had a definite balance form. The reason of the
difference is hidden in the way of the generalization of the entropy
current. The classical form reads as a product of the currents of the
extensives and the corresponding entropic intensives
\begin{equation}
 {\bf j}_s = \partial_a s {\bf j}_a
\label{ClassEntCurr}\end{equation}

In the generalization of Ny\'\i{}ri, in section \ref{NEITex} developing
weakly nonlocal extended irreversible thermodynamics, the entropy
current is assumed to be a homogeneous linear function of the currents
of the extensives
\begin{equation}
 {\bf j}_s = {\bf B}\cdot {\bf j}_a.
\label{NyirEntCurr}\end{equation}

On the other hand, in case of dynamic variables that vanish in
equilibrium and the entropy characterizes the equilibrium the entropy
current was assumed to be a linear function of the entropic intensives,
conjugated to the corresponding extensive variable.
\begin{equation}
{\bf j}_s = {\bf A}\cdot \partial_\xi s.
\label{BalEntCurr}\end{equation}

This form is a different generalization of (\ref{ClassEntCurr}) than
(\ref{NyirEntCurr}) and was applied in section \ref{NEITex} related to
model B, leading to the thermodynamic Cahn-Hilliard equation. Extended
thermodynamics looks like a special case of the system treated in
\ref{NEITex}, because currents of extensives can be considered as
special internal variables. How the two assumptions are related to each
other? Could we arrive to an evolution equation with balance form also
in case of extended thermodynamics as it is generally expected
\cite{MulRug98b,JouAta92b}? To answer this question we should
investigate the forms of the entropy inequality more closely. Instead
of simply refering to the results of the previous sections we will give
again a less refined, briefer, more direct, althought sloppier,
traditional irreversible thermodynamic approach.

To get the entropy production of weakly nonlocal extended
thermodynamics (\ref{EntrProdWNExt}) we consider a system with a basic
state spanned by extensive {\bf a} and internal variable
$\boldsymbol{\xi}$ together. In case of extended thermodynamics the
(dynamic) internal variable is the (dynamic) current of the extensives
$\boldsymbol{\xi} = {\bf j}_a$. Assuming a local entropy function
$s({\bf a}, {\bf j}_a)$ one can get (\ref{EntrProdWNExt}) by
substituting (\ref{NyirEntCurr}), (\ref{EvJ}) and (\ref{BalExt1}) into
the entropy balance as follows
\begin{gather}
\partial_t s + \nabla\cdot {\bf j}_s = \nonumber\\
\partial_as \cdot \partial_t {\bf a} +
    \partial_j s \cdot \partial_t {\bf j}_a +
    \nabla\cdot ({\bf B}\cdot {\bf j}_a) =  \nonumber\\
-\partial_as \cdot \nabla\cdot{\bf j}_a -
    \partial_j s \cdot \mathcal{G} +
    {\bf B}:\nabla {\bf j}_a +
    \nabla\cdot{\bf B}\cdot {\bf j}_a = \nonumber\\
({\bf B} - \partial_as I): \nabla{\bf j}_a  +
    (\nabla\cdot {\bf B} + \mathcal{G}\widehat{\bf m})\cdot {\bf j}_a
\geq 0.
\label{IrrevWNExt}\end{gather}

Here we exploited the Gyarmati form entropy (\ref{WaveEntropy}) and
wrote $\partial_js = -\widehat{\bf m} \cdot {\bf j}_a$. Because the
entropy function is the primary constitutive quantity we can conclude
that (\ref{IrrevWNExt}) resulted in a solvable inequality for he two
(!) constitutive functions ${\bf B}$ and $\mathcal{G}$.

On the other hand, when the basic state space is the same as previously
but with a general dynamic internal variable one evidently cannot
assume any more that the entropy current is proportional to the dynamic
variable and one cannot write the entropy current in a one term form of
(\ref{BalEntCurr}). However, a splitting into two terms, a classical
and a non classical one, works well. Therefore we will assume the
following configuration of the entropy current
\begin{equation}
{\bf j}_s = \partial_as {\bf j}_a + {\bf B}'\cdot\boldsymbol{\xi}.
\end{equation}

In this case a we can go forward similarly to the previous derivation
and substitute (\ref{BalExt1}) and Gyarmati entropy into the balance of
entropy to get
\begin{gather}
\partial_t s + \nabla\cdot {\bf j}_s = \nonumber\\
\partial_as \cdot \partial_t {\bf a} +
    \partial_\xi s \cdot \partial_t \boldsymbol{\xi} +
    \nabla\cdot (\partial_as {\bf j}_a +
        {\bf B}'\cdot \boldsymbol{\xi}) = \nonumber\\
-\partial_as \cdot \nabla\cdot{\bf j}_a -
    \partial_\xi s \cdot \mathcal{F} +
    \partial_as \nabla\cdot {\bf j}_a +
    {\bf j}_a\cdot \nabla \partial_as  +
    {\bf B}':\nabla \boldsymbol{\xi} +
    \nabla\cdot{\bf B}'\cdot \boldsymbol{\xi} = \nonumber\\
{\bf B}': \nabla\boldsymbol{\xi}  +
    {\bf j}_a\cdot \nabla \partial_as  +
    (\nabla\cdot {\bf B}' + \widehat{\bf m} \mathcal{F})\cdot
    \boldsymbol{\xi}.
\geq 0. \label{IrrevWNCH1}\end{gather}

This a solvable inequality for he three (!) constitutive functions
${\bf B}'$, $\mathcal{F}$ and ${\bf j}_a$. Moreover it reduces to the
previous (\ref{IrrevWNExt}) one in the special case of
$\boldsymbol{\xi} = {\bf j}_a$, when $\mathcal{G} = \mathcal{F}$ and
${\bf B}= {\bf B}' + \partial_as I$. Therefore  ${\bf j}_s = ({\bf B}'
+ \partial_as I)\cdot {\bf j}_a$. However, the solution of the
inequality does not result in a balance form evolution equation for the
internal variable, similarly to the extended thermodynamic case.

In section \ref{CHex} the entropy current was assumed to have the
following, different form
\begin{equation}
{\bf j}_s = \partial_as {\bf j}_a + {\bf A}\cdot \partial_\xi s.
\label{sjch}\end{equation}

Now, we can substitute (\ref{BalExt1}) and (\ref{CHEntCurr}) into the
balance of entropy to get the entropy inequality in the following
configuration
\begin{gather}
\partial_t s + \nabla\cdot {\bf j}_s = \nonumber\\
\partial_as \cdot \partial_t {\bf a} +
    \partial_\xi s \cdot \partial_t \boldsymbol{\xi} +
    \nabla\cdot (\partial_as {\bf j}_a +
        {\bf A}\cdot \partial_\xi s) =  \nonumber\\
-\partial_as \cdot \nabla\cdot{\bf j}_a -
    \partial_\xi s \cdot \mathcal{F} +
    \partial_as \nabla\cdot {\bf j}_a +
    {\bf j}_a\cdot \nabla \partial_as  +
    {\bf A}:\nabla \partial_\xi s +
    \nabla\cdot{\bf A}\cdot \partial_\xi s = \nonumber\\
{\bf A}: \nabla\partial_\xi s  +
    {\bf j}_a\cdot \nabla \partial_as  +
    (\nabla\cdot {\bf A} - \mathcal{F})\cdot
    \partial_\xi s.
\geq 0. \label{IrrevWNCH2}\end{gather}

This a solvable inequality for he three (!) constitutive functions
${\bf A}$, $\mathcal{F}$ and ${\bf j}_a$ again. Moreover, as we have
seen in section \ref{CHex} the solution results in a balance form
evolution equation of the internal variable without applying Gyarmati
entropy.

In the special case of extended thermodynamics, when $\boldsymbol{\xi}
= {\bf j}_a$ the entropy inequality transforms to
$$
\sigma_s = {\bf A}: \nabla\partial_j s  +
    {\bf j}_a\cdot \nabla \partial_as  +
    (\nabla\cdot {\bf A} - \mathcal{F})\cdot
    \partial_j s \geq 0.
$$

As one can see the above inequality does not have a general solution
being a sum of three product terms with two constitutive functions.
However, with Gyarmati entropy $\partial_j s = -\widehat{\bf m} \cdot
{\bf j}_a$ and the inequality further transforms to
$$
\sigma_s = \widehat{\bf m}\cdot {\bf A} \cdot \nabla{\bf j}_a  +
    \left(\nabla\cdot (\partial_as I - \widehat{\bf m}\cdot{\bf A}) -
     \widehat{\bf m}\cdot\mathcal{F}\right) \cdot {\bf j}_a \geq 0.
$$

Here we can recognize the classical solvable extended thermodynamic
form with ${\bf B}' = \widehat{\bf m}\cdot {\bf A}$ and the entropy
current in the Ny\'\i{}ri form as ${\bf j}_s = {\bf B}\cdot {\bf j}_a =
(-\widehat{\bf m}\cdot {\bf A} + \partial_as I)\cdot{\bf j}_a$. The
inequality was transformed into a solvable force-current system, but
the balance form of the arising evolution equation of the internal
variable is destroyed.

As a summary of the previous investigations we have seen that in case
of dynamic internal variables in general the balance form of the
evolution equation is a consequence of natural physical conditions but
in case of extended thermodynamics the limited constitutive structure
of the theory seem to destroy the balance form of the evolution
equations as general solutions of the dissipation inequality.

Let us remark here, that the easier formal applicability and heuristic
power of the classical irreversible thermodynamic method can support
but cannot replace the deeper constitutive investigations of the first
section. First of all because the postulates of the more exact method
were clearer and weaker than the heuristic one (e.g. the form of the
entropy current (\ref{CHEntCurr}) was consequence of the Liu
equations). On the other hand one should identify correctly the basic
state space and the constitutive space to see beyond the rather formal
treatment of current-force systems of irreversible thermodynamics.

\section{Summary, conclusions and discussion}

In the previous sections we have seen that it is possible to extend
irreversible thermodynamics to nonlocal phenomena with the help of
generalized internal variables. The extension is based on the
generalization of the entropy current. Partial differential equations
were constructed constitutively as evolution equations of the internal
variables by a force-current structure. The structure of
nonlinearities, the sign of coefficients were determined by the
requirement of nonnegative entropy production. We have investigated the
thermodynamic background of several classical equations of mathematical
physics containing higher order space derivatives.

In the first section we have seen that the choice of the variables,
basic and constitutive state spaces are crucial steps in thermodynamic
theory construction. Basic state spaces are spanned by the physical
quantities for that one wants to find an evolution equation.
Constitutive state spaces are spanned by variables of the basic state
space and their derivatives, the ones, the constitutive functions are
supposed to depend on. The extension of the constitutive state space
determines whether an evolution equation of an extensive or an internal
variable considers nonlocal (or memory) effects or not.

In this paper we investigated the nonlocal extension only of dynamic
internal variables. According to this view we can call a basic internal
state variable nonlocal if the constitutive space contains at least its
second order space derivatives. The following table summarizes the
hierarchy of continuum theories from that point of view. {\bf a}
denotes the array of extensive variables and $\boldsymbol{\xi}$ an
array of internal variables as above.

\vskip 0.1in
\setlength{\extrarowheight}{2mm}
\begin{tabular}{|p{1.2in}|p{0.7in}||p{1in}|p{1.3in}|}\hline
    & Basic state &
    \multicolumn{2}{c}{{Constitutive state}} \vline \\ \hline\hline
    & &  Local &  Nonlocal \\\hline
Classical
    & {\bf a}
    & $({\bf a}, \nabla{\bf a})$
    & \\\hline
Extended
    & $({\bf a}, {\bf j}_a)$
    & $({\bf a}, {\bf j}_a, \nabla{\bf a})$
        & $({\bf a}, {\bf j}_a, \nabla{\bf a}, \nabla{\bf j}_a,
            \nabla^2{\bf j}_a)$ \\\hline
Internal variable
    & $\boldsymbol{\xi}$
    & $(\boldsymbol{\xi}, \nabla\boldsymbol{\xi})$
        & $(\boldsymbol{\xi}, \nabla\boldsymbol{\xi},
            \nabla^2\boldsymbol{\xi})$ \\\hline
Classical + internal variable
        & $({\bf a},\boldsymbol{\xi})$
        & $({\bf a},\boldsymbol{\xi},\nabla{\bf a})$
        & $({\bf a}, \boldsymbol{\xi}, \nabla{\bf a},
            \nabla\boldsymbol{\xi}, \nabla^2\boldsymbol{\xi})$ \\\hline
\end{tabular}
\vskip 0.1in

\begin{itemize}
\item The first row of the table concerns classical irreversible
thermodynamics where the basic state space is spanned by specific
extensive quantities. Classical irreversible thermodynamics is a local
theory, in the sense of local equilibrium. Investigating its nonlocal
extensions one can easily conclude that in case of relocalizable
theories (when entropy depends only on the basic state), the structure
of the equations does not change and only the classical constitutive
functions become gradient dependent.
\item In the second row, where the basic state space is spanned by the
extensive variables ${\bf a}$ and their (dynamic) current densities
${\bf j}_a$ one get classical extended thermodynamics. Extending the
constitutive space with the gradients of the current densities gives a
nonlocal theory and leads to Guyer-Krumhansl-like equations.
 \item For pure weakly nonlocal dynamic internal variables
$\boldsymbol{\xi}$ the nonlocal extension of the classical ordinary
differential equations results in Ginzburg-Landau-like equations.
 \item An extensive variable with a weakly nonlocal dynamic internal
variable leads to Cahn-Hilliard-like equations.
\end{itemize}

Our investigations were restricted to relocalizable theories, where the
nonlocality can be grasped through nonclassical internal variables,
introduced into the entropy current. The examples can be continued,
other kind of basic and constitutive states and couplings can lead to
other weakly nonlocal thermodynamic compatible equations. However, we
may not forget that there are genuine weakly nonlocal theories and
equations, with large importance and ample experimental evidence. A
typical example is the traditional, non thermodynamic  Ginzburg-Landau
equation. Fortunately our methods are applicable to embed also true
weakly nonlocal equations into nonequilibrium thermodynamics. The key
aspects of Liu's procedure regarding that question and the first
investigations related to the compatibility of the Ginzburg-Landau
equation with nonequilibrium thermodynamics are given in \cite{Van02m}.

We can observe that the emerging classification scheme is more refined
that was given by Penrose and Five \cite{PenFif90a}, Hohenberg and
Halperin \cite{HohHal77a} or Gurtin \cite{Gur96a}. The investigation of
several existing weakly nonlocal equations (e.g. different phase field
equations, the complex Ginzburg-Landau or Kardar-Parisi-Zhang) from the
point of view of nonequilibrium thermodynamics can reveal richer
structures, new terms, clearer conditions, surprising interrelations,
hence a  broader range of applicability than usually considered.

In this paper it was shown that contrary to the common belief,
irreversible thermodynamics can be extended far beyond the usual local
equilibrium or local state hypotheses, one can consider dissipation
related to nonlocal effects, too. A rigorous treatment showed that the
seemingly intuitive steps of irreversible thermodynamic modelling are
well supported and explained with Liu procedure. Several interesting
consequences, new terms in the traditional equations and possible
sources of physical generalizations were explored. There are several
experimental and theoretical evidences that the characteristic
additional term, containing the time derivative of the Laplacian of the
basic state variable ($\Delta\partial_t$), appears in all of the
treated classical equations (Guyer-Krumhansl, Ginzburg-Landau and
Cahn-Hilliard). That term cannot be derived from variational
considerations without any further ado, because its Fre'chet (strong)
derivative is not symmetric \cite{VanMus95a}.

An other important remark, that the stability structure of
nonequilibrium thermodynamics is preserved in the weakly nonlocal
extension. Therefore the entropy can serve as an Liapunov functional
with appropriate initial and boundary conditions. These possible
boundary conditions are more general than that were given by Penrose
and Fife in their investigations regarding the thermodynamic
compatibility of Ginzburg-Landau and Cahn-Hilliard equations. The
physical difference is that now Liapunov functions can be generated for
several different open systems. This kind of generalization goes
farther than it was possible in homogeneous systems (see e.g.
\cite{Mat00a}) and also requires a specification of the function spaces
where the solutions of the differential equations are investigated.

\section{Acknowledgements}

Thanks for Christina Papenfuss and Tam\'as Matolcsi for careful
reading of the manuscript, Antonio Cimmelli and P\'eter B. B\'eda for
their valuable remarks. This research was supported by OTKA T034715 and
T034603.

\section{Appendix A: Liu-procedure}

The Liu-procedure is based on an application of a theorem in
linear algebra. That theorem can be considered as a consequence of
the Farkas' lemma \cite{Far894a}, well known in linear
programming. More properly it is a special case of the so called
affine Farkas' lemma \cite{Haa18a} (that can be easily derived
from the original Farkas' lemma \cite{Far18a2}). The recognition
of this connection between the Liu theorem and Farkas lemma is due
to some recent researches of Kirchner and Hauser \cite{HauKir02a}.

In this appendix we formulate and prove that theorem of Liu
\cite{Liu72a} in a simple and general form best suited for our
purpose. The proof below is essentially that of Liu.

$\mathbb{ V}$ and $\mathbb{ V'}$ are finite dimensional vector
spaces, $\mathbb{ V}^*$ and $\mathbb{ V'}^*$ denote their duals.

\begin{thm}(Liu)
${\bf A} \in Lin(\mathbb{V},\mathbb{ V'})$, ${\bf B} \in \mathbb{
V'}$, ${\bf a} \in \mathbb{ V}^*$ and $b \in \mathbb{R}$. Then
${\bf a}\cdot {\bf x} + b \geq 0$ holds for all ${\bf x}\in
\mathbb{ V}$ such that ${\bf A}\cdot{\bf x} + {\bf B} = {\bf
0}_\mathbb{ V'}$ if and only if there is a $\lambda \in \mathbb{
V'}^*$ such that
\begin{eqnarray}
{\bf a} - \lambda\cdot {\bf A} &=&{\bf 0}_\mathbb{ V^*}, \label{Liu-eq}\\
b-\lambda\cdot {\bf B} &\geq& 0. \label{Dis-ineq}
\end{eqnarray}

(\ref{Liu-eq}) is called {\em Liu-equation} and (\ref{Dis-ineq})
is the {\em dissipation inequality} in the thermodynamic
literature. \label{Liu-th}\end{thm}

Proof:

The backward direction is easier. If there is a $\lambda$ with the
above property, then for all ${\bf x}\in \mathbb{ V}$, $ 0 \leq
({\bf a} - \lambda\cdot{\bf A})\cdot {\bf x} - \lambda\cdot{\bf B}
+ b = ({\bf a}\cdot{\bf x} + b)  - \lambda\cdot({\bf A}\cdot {\bf
x} + {\bf B})$ and the statement follows.

Now let us prove the forward direction. We introduce the notation
$\mathcal{A}ff= \left\{ {\bf x}\in \mathbb{ V} \left|\right. {\bf
A}\cdot{\bf x} + {\bf B} = {\bf 0}_\mathbb{ V'}\right\}$ . First of all
we should see that $Ker {\bf A} \subset Ker {\bf a}$. Really if ${\bf
y}\in Ker{\bf A}$ and ${\bf x} \in \mathcal{A}ff$ then for all $r\in
\mathbb{R}$, ${\bf x} + r {\bf y} \in \mathcal{A}ff$, therefore
according to the conditions of the theorem $ 0\leq {\bf a}\cdot({\bf x}
+ r {\bf y}) + b = {\bf a\cdot x}+b + r {\bf a\cdot y}. $ For arbitrary
$r$ the inequality is true if and only if ${\bf a\cdot y}=0$ therefore
$Ker {\bf A} \subset Ker {\bf a}$.

For the annullators of the kernels
$$
(Ker {\bf a})^\perp \subset (Ker {\bf A})^\perp,
$$

\noindent because $\forall {\bf x} \in Ker
{\bf A} \subset Ker {\bf a}$, therefore, if ${\bf p}\in(Ker {\bf
a})^\perp$ then ${\bf p}\cdot {\bf x} = 0$.

On the other hand $Ker {\bf A}^* = (Ran {\bf A})^\perp$, because if
$\forall {\bf x}\in \mathbb{ V} \quad {\bf y\cdot A \cdot x} = 0 = {\bf
x \cdot A^* \cdot y}$ then from the left hand side of the equality
${\bf y} \in (Ran {\bf A})^\perp$ and from the right hand side ${\bf
y}\in Ker {\bf A}^*$. For finite dimensional $\mathbb{ V}, {\bf A}^{**}
= {\bf A}$ and $({\bf A}^\perp)^\perp = {\bf A}$, therefore we can
transform the equality into the form
$$
(Ker {\bf A})^\perp = Ran {\bf A}^*.
$$
Putting together all these statements we can see, that ${\bf a}\in
(Ker {\bf a})^\perp \subset (Ker{\bf A})^\perp = Ran {\bf A}^*$,
therefore there is a $\lambda \in {\mathbb{ V}'}^*$ such that
${\bf a}={\bf A}^*\cdot \lambda$. In finite dimension follows
(\ref{Liu-eq}). Therefore we can write for all ${\bf x}$ with
${\bf A}\cdot{\bf x} +{\bf B} ={\bf 0}_\mathbb{ V'}$ that $0 \leq
{\bf a}\cdot {\bf x} + b = \lambda\cdot{\bf A}\cdot {\bf x} + b =
b - \lambda\cdot{\bf B}. $

The proof is complete. \hfill

\begin{rem}
If {\bf A} is surjective ($dim(Rank {\bf A}) = dim \mathbb{ V}'$)
then $\lambda$ is unique.
\end{rem}

%Possible generalizations:
%1 with pure affine maps -> impossible
%2 with more inequalities: a\in Lin(V,V') and dim(V')\noneq 1

\section{Appendix B: Mean-value theorem}

We applied Lagrange's mean value theorem several times in the
paper. Here we formulate it in a way that is sufficiently general
for our purpose. One can find a proof for example in \cite{Rud70b}.

\begin{thm}(Lagrange)
Let $\mathbb{ X}$ be a normed space. $F:\mathbb{ X} \rightarrow
\mathbb{R}$, ${\bf x}, {\bf y}\in \mathbb{ X}$. If $F$ is
differentiable on the closed section $[{\bf x},{\bf y}]$, then
there is a ${\bf z} \in [{\bf x},{\bf y}]$, so that
$$
F({\bf y}) - F({\bf x}) = DF({\bf z})\cdot ({\bf y}-{\bf x}),
$$

\noindent where $DF({\bf z})$ denotes the derivative of $F$ at
${\bf z}$.
\end{thm}

\section{Appendix C: Solution of the Dissipation Inequality}

In a force-current form the dissipation inequality can be written
as a sum of the product of constitutive quantities and variables
in the constitutive space. In this case a solution can be given
in the form of the (nonlinear) Onsagerian conductivity relations.
we formulate the statement and give the simple proof following
Gurtin \cite{Gur96a}.

\begin{thm}
Let $\mathbb{ V}$ be a finite dimensional vector space. ${\bf
J}:\mathbb{ V} \rightarrow \mathbb{ V}^*$, ${\bf x} \mapsto {\bf
J}({\bf x})$ is a continuously differentiable function. If ${\bf
J}({\bf x})\cdot {\bf x} \geq 0$ for all ${\bf x}\in \mathbb{ V}$
then ${\bf J}({\bf x}) = {\bf L}({\bf x}){\bf x}$ where ${\bf L}:
\mathbb{ V} \rightarrow Bilin(\mathbb{ V})$.
\end{thm}

Proof:

From the inequality follows that for all positive number $\lambda$
and ${\bf x}\in \mathbb{ V}$, ${\bf J}(\lambda {\bf
x})\cdot\lambda{\bf x} \geq 0$, therefore ${\bf J}(\lambda {\bf
x})\cdot {\bf x}\geq 0$. In the limit when $\lambda $ goes to zero
we get that ${\bf J}({\bf 0}) \cdot {\bf x} \geq 0$. That can be
true for all ${\bf x}\in U$ if and only if ${\bf J}({\bf 0}) =
{\bf 0}$. The statement follows according to the mean value
theorem.

\begin{rem}
The structure is not unique in case of nonlinear relations.
\end{rem}


\begin{thebibliography}{10}

\bibitem{Rog79a}
D.~Rogula.
\newblock Geometrical and dynamical nonlocality.
\newblock {\em Archives of Mechanics (Stosowanej)}, 31(1):65--75, 1979.

\bibitem{Eri99b}
C.~Eringen.
\newblock {\em Microcontinuum Field Theories I. Foundations and Solids}.
\newblock Springer-Verlag, Berlin-etc.., 3th edition, 1999.

\bibitem{Ede74a}
D.~G.~B. Edelen.
\newblock Irreversible thermodynamics of nonlocal systems.
\newblock {\em International Journal of Engineering Science}, 12:607--631,
  1974.

\bibitem{Rog82c}
D.~Rogula.
\newblock Introduction to nonlocal theory of material media.
\newblock In D.~Rogula, editor, {\em Nonlocal Theory of Material Media}, volume
  268 of {\em CISM Courses and Lectures}, pages 123--222. Springer Verlag,
  Wien-New York, 1982.

\bibitem{Mau79a}
G.~A. Maugin.
\newblock Nonlocal theories or gradient-type theories: a matter of convenience?
\newblock {\em Archives of Mechanics (Stosowanej)}, 31(1):15--26, 1979.

\bibitem{TruNol65b}
C.~Truesdell and W.~Noll.
\newblock {\em The Non-Linear Field Theories of Mechanics}.
\newblock Springer Verlag, Berlin-Heidelberg-New York, 1965.
\newblock Handbuch der Physik, III/3.

\bibitem{Ver97b}
J.~Verh\'as.
\newblock {\em Thermodynamics and {R}heology}.
\newblock Akad\'emiai Kiad\'o and Kluwer Academic Publisher, Budapest, 1997.

\bibitem{KosWoj95a}
W.~Kosi\'nski and W.~Wojno.
\newblock A gradient generalization to internal state variable approach.
\newblock {\em Archive of Mechanics}, 47(3):523--536, 1995.

\bibitem{Kos96p}
W.~Kosi\'nski.
\newblock A modified hyperbolic framwork for thermoelastic materials with
  damage.
\newblock In W.~Kosi\'nski, R.~de~Boer, and D.~Gross, editors, {\em Problems of
  Environmental and Damage Mechanics}, pages 157--172, Warszawa, 1997.
  IPPT-PAN.
\newblock Proceedings of SolMech'96 Conference, Mierki, Sept. 9-14. 1996,
  Poland.

\bibitem{CimKos97a}
V.~A. Cimmelli and W.~Kosi\'nski.
\newblock Gradient generalization to internal state variables and a theory of
  superfluidity.
\newblock {\em Journal of Theoretical and Applied Mechanics}, 35(4):763--779,
  1997.

\bibitem{Val96a}
K.~C. Valanis.
\newblock A gradient theory of internal variables.
\newblock {\em Acta Mechanica}, 116:1--14, 1996.

\bibitem{Val98a}
K.~C. Valanis.
\newblock A gradient thermodynamic theory of self-organization.
\newblock {\em Acta Mechanica}, 127:1--23, 1998.

\bibitem{VarAif94a}
I.~Vardoulakis and E.~C. Aifantis.
\newblock On the role of microstructure in behaviour of solids: effects of
  higher order gradients and internal inertia.
\newblock {\em Mechanics of Materials}, 18:151--158, 1994.

\bibitem{Bed00a}
P.~B. B\'eda.
\newblock Dynamic systems, rate and gradient effects in material instability.
\newblock {\em International Journal of Mechanical Sciences}, 42:2101--2114,
  2000.

\bibitem{Mau90a1}
G.~A. Maugin.
\newblock Internal variables and dissipative structures.
\newblock {\em Journal of Non-Equilibrium Thermodynamics}, 15:173--192, 1990.

\bibitem{Mau80a}
G.~A. Maugin.
\newblock The principle of virtual power in continuum mechanics. {A}pplication
  to coupled fields.
\newblock {\em Acta Mechanica}, 35:1--70, 1980.

\bibitem{CapMar01a}
G.~Capriz and P.~M. Mariano.
\newblock Multifield theories: and introduction.
\newblock {\em International Journal of Solids and Structures}, 38:939--941,
  2001.
\newblock Preface to a special issue dedicated to multifield theories.

\bibitem{Mar02a}
P.~M. Mariano.
\newblock Multifield theories in mechanics of solids.
\newblock {\em Advances in Applied Mechanics}, 38:1--94, 2002.

\bibitem{GroMaz62b}
S.~R. de~Groot and P.~Mazur.
\newblock {\em Non-equilibrium Thermodynamics}.
\newblock North-Holland Publishing Company, Amsterdam, 1962.

\bibitem{Gya70b}
I.~Gyarmati.
\newblock {\em Non-equilibrium Thermodynamics /{F}ield Theory and Variational
  Principles/}.
\newblock Springer Verlag, 1970.

\bibitem{ColNol63a}
B.~D. Coleman and W.~Noll.
\newblock The thermodynamics of elastic materials with heat conduction and
  viscosity.
\newblock {\em Archive for Rational Mechanics and Analysis}, 13:167--178, 1963.

\bibitem{Liu72a}
I-Shih Liu.
\newblock Method of {L}agrange multipliers for exploitation of the entropy
  principle.
\newblock {\em Archive of Rational Mechanics and Analysis}, 46:131--148, 1972.

\bibitem{Hut77a}
K.~Hutter.
\newblock The foundations of thermodynamics, its basic postulates and
  implications. {A} review of modern thermodynamics.
\newblock {\em Acta Mechanica}, 27:1--54, 1977.

\bibitem{Sil97b}
M.~$\check{S}$ilhav\'y.
\newblock {\em The Mechanics and Thermodynamics of Continuous Media}.
\newblock Springer Verlag, Berlin-etc., 1997.

\bibitem{MusEhr96a}
W.~Muschik and H.~Ehrentraut.
\newblock An amendment to the {S}econd {L}aw.
\newblock {\em Journal of Non-Equilibrium Thermodynamics}, 21:175--192, 1996.

\bibitem{Mat92a1}
T.~Matolcsi.
\newblock Dynamical laws in thermodynamics.
\newblock {\em Physics Essays}, 5(3):320--327, 1992.

\bibitem{GlaPri71b}
P.~Glansdorff and I.~Prigogine.
\newblock {\em Thermodynamic Theory of Structure, Stability and Fluctuations}.
\newblock Wiley-Interscience, London-etc., 1971.

\bibitem{Kes90a1}
J.~Kestin.
\newblock A note on the relation between the hypothesis of local equilibrium
  and the {C}lausius-{D}uhem inequality.
\newblock {\em Journal of Non-Equilibrium Thermodynamics}, 15:193--212, 1990.

\bibitem{Kes93a1}
J.~Kestin.
\newblock Internal variables in the local-equilibrium approximation.
\newblock {\em Journal of Non-Equilibrium Thermodynamics}, 18:360--379, 1993.

\bibitem{OnsMac53a}
L.~Onsager and S.~Machlup.
\newblock Fluctuations and irreversible processes.
\newblock {\em Physical Reviews}, 91:1505--1512, 1953.

\bibitem{MacOns53a}
S.~Machlup and L.~Onsager.
\newblock Fluctuations and irreversible process. {II}. {S}ystems with kinetic
  energy.
\newblock {\em Physical Reviews}, 91:1512--1515, 1953.

\bibitem{Mat95a}
T.~Matolcsi.
\newblock Reservoirs in thermodynamics.
\newblock {\em Physics Essays}, 8(3):234--239, 1995.

\bibitem{Van95a}
P.~V\'an.
\newblock Another {D}ynamic {L}aws in thermodynamics.
\newblock {\em Physics Essays}, 8(4):457--465, 1995.

\bibitem{Mat96a1}
T.~Matolcsi.
\newblock On the classification of phase transitions.
\newblock {\em ZAMP}, 47(6):837--857, 1996.

\bibitem{MusDom96a}
W.~Muschik and R.~Dom\'\i{}nguez-Cascante.
\newblock On extended thermodynamics of discrete systems.
\newblock {\em Physica A}, 233:523--550, 1996.

\bibitem{Mat96a2}
T.~Matolcsi.
\newblock On the dynamics of phase transitions.
\newblock {\em ZAMP}, 47(6):858--879, 1996.

\bibitem{Mat00a}
T.~Matolcsi.
\newblock On the mathematical structure of thermodynamics.
\newblock {\em Journal of Mathematical Physics}, 41(4):2021--2042, 2000.

\bibitem{Mul68a}
I.~M\"uller.
\newblock A thermodynamic theory of mixtures of fluids.
\newblock {\em Archive of Rational Mechanics and Analysis}, 28:1--39, 1968.

\bibitem{MulRug98b}
I.~M\"uller and T.~Ruggeri.
\newblock {\em Rational Extended Thermodynamics}, volume~37 of {\em Springer
  Tracts in Natural Philosophy}.
\newblock Springer Verlag, New York-etc., 2nd edition, 1998.

\bibitem{Ver83a}
J.~Verh\'as.
\newblock On the entropy current.
\newblock {\em Journal of Non-Equilibrium Thermodynamics}, 8:201--206, 1983.

\bibitem{LebAta97a}
G.~Lebon, D.~Jou, J.~Casas-V\'azquez, and W.~Muschik.
\newblock Heat conduction at low temperature: A non-linear generalization of
  the {G}uyer-{K}rumhansl equation.
\newblock {\em Periodica Polytechnica Chemical Engineering}, 41(2):185--196,
  1997.
\newblock Lecture held on 'Minisymposium on Non-Linear Thermodynamics and
  Reciprocal Relations', September 22-26, Balatonvil\'agos, Hungary.

\bibitem{MarAug98a}
P.~M. Mariano and G.~Augusti.
\newblock Multifield description of microcracked continua: A local model.
\newblock {\em Mathematics and Mechanics of Solids}, 3:183--200, 1998.

\bibitem{MusAta01a}
W.~Muschik, C.~Papenfuss, and H.~Ehrentraut.
\newblock A sketch of continuum thermodynamics.
\newblock {\em Journal of Non-Newtonian Fluid Mechanics}, 96:255--290, 2001.

\bibitem{Mat84b}
T.~Matolcsi.
\newblock {\em A Concept of Mathematical Physics: Models for SpaceTime}.
\newblock Akad\'emiai Kiad\'o (Publishing House of the Hungarian Academy of
  Sciences), Budapest, 1984.

\bibitem{Mat85a}
T.~Matolcsi.
\newblock On material frame-indifference.
\newblock {\em Archive of Rational Mechanics and Analysis}, 91(2):99--118,
  1986.

\bibitem{BalVan02m}
M.~Bal\'azs and P.~V\'an.
\newblock Lagrange formalism of point-masses.
\newblock mat-ph/0205040, 2002.

\bibitem{Gya77a}
I.~Gyarmati.
\newblock The wave approach of thermodynamics and some problems of non-linear
  theories.
\newblock {\em Journal of Non-Equilibrium Thermodynamics}, 2:233--260, 1977.

\bibitem{Gya69a}
I.~Gyarmati.
\newblock On the {G}overning {P}rinciple of {D}issipative {P}rocesses and its
  extension to nonlinear problems.
\newblock {\em Annalen der Physik}, 23:353--378, 1969.

\bibitem{Van96a}
P.~V\'an.
\newblock On the structure of the '{G}overning {P}rinciple of {D}issipative
  {P}rocesses'.
\newblock {\em Journal of Non-Equilibrium Thermodynamics}, 21(1):17--29, 1996.

\bibitem{JouAta92b}
D.~Jou, J.~Casas-V\'azquez, and G.~Lebon.
\newblock {\em Extended Irreversible Thermodynamics}.
\newblock Springer Verlag, Berlin-etc., 1992.
\newblock 3rd, revised edition, 2001.

\bibitem{JouAta99a}
D.~Jou, J.~Casas-V\'azquez, and G.~Lebon.
\newblock Extended irreversible thermodynamics revisited (1988-98).
\newblock {\em Reports on Progress in Physics}, 62:1035--1142, 1999.

\bibitem{Gil81b}
R.~Gilmore.
\newblock {\em Catastrophe Theory for Scientists and Engineers}.
\newblock John Wiley and Sons, New York, etc., 1981.

\bibitem{Nyi91a1}
B.~Ny\'\i{}ri.
\newblock On the entropy current.
\newblock {\em Journal of Non-Equilibrium Thermodynamics}, 16:179--186, 1991.

\bibitem{MauMus94a1}
G.~A. Maugin and W.~Muschik.
\newblock Thermodynamics with internal variables. {P}art {I}. {G}eneral
  concepts.
\newblock {\em Journal of Non-Equilibrium Thermodynamics}, 19:217--249, 1994.

\bibitem{MauMus94a2}
G.~A. Maugin and W.~Muschik.
\newblock Thermodynamics with internal variables. {P}art {II}. {A}pplications.
\newblock {\em Journal of Non-Equilibrium Thermodynamics}, 19:250--289, 1994.

\bibitem{ColGur67a}
B.~D. Coleman and M.~E. Gurtin.
\newblock Thermodynamics with internal state variables.
\newblock {\em The Journal of Chemical Physics}, 47(2):597--613, 1967.

\bibitem{OrlRoz84a1}
V.~N. Orlov and L.~I. Rozonoer.
\newblock The macrodynamics of open systems and the variational principle of
  the local potential -- {I}.
\newblock {\em Journal of Franklin Institute}, 318:283--314, 1984.

\bibitem{OrlRoz84a2}
V.~N. Orlov and L.~I. Rozonoer.
\newblock The macrodynamics of open systems and the variational principle of
  the local potential -- {II} {A}pplications.
\newblock {\em Journal of Franklin Institute}, 318:315--347, 1984.

\bibitem{Fal92a}
F.~Falk.
\newblock {C}han-{H}illiard theory and irreversible thermodynamics.
\newblock {\em Journal of Non-Equilibrium Thermodynamics}, 17:53--65, 1992.

\bibitem{ZbiAif88a1}
H.~M. Zbib and E.~C. Aifantis.
\newblock On the localization an postlocalization behaviour of plastic
  deformation. {I}. {O}n the initiation of shear bands.
\newblock {\em Res Mechanica}, 23:261--277, 1988.

\bibitem{HohHal77a}
P.~C. Hohenberg and B.~I. Halperin.
\newblock Theory of dynamic critical phenomena.
\newblock {\em Reviews of Modern Physics}, 49(3):435--479, 1977.

\bibitem{GrmOtt97a}
M.~Grmela and H.~C. \"Ottinger.
\newblock Dynamics and thermodynamics of complex fluids. {I}. {D}evelopment of
  a general formalism.
\newblock {\em Physical Review E}, 56(6):6620--6632, 1997.

\bibitem{OttGrm97a}
H.~C. \"Ottinger and M.~Grmela.
\newblock Dynamics and thermodynamics of complex fluids. {II}. {I}llustrations
  of a general formalism.
\newblock {\em Physical Review E}, 56(6):6633--6655, 1997.

\bibitem{Gur96a}
M.~G. Gurtin.
\newblock Generalized {G}inzburg-{L}andau and {C}ahn-{H}illiard equations based
  on a microforce balance.
\newblock {\em Physica D}, 92:178--192, 1996.

\bibitem{Van02a1}
P.~V\'an.
\newblock Weakly nonlocal irreversible thermodynamics - the {G}inzburg
  -{L}andau equation.
\newblock {\em Technische Mechanik}, 22(2):104--110, 2002.
\newblock cond-mat/0111307.

\bibitem{Ver96a}
J.~Verh\'as.
\newblock Once again on the transport of dynamic degrees of freedom.
\newblock {\em Atti Accademia Peloritana dei Pericolanti}, LXXII:101--114,
  1996.

\bibitem{Van01a1}
P.~V\'an.
\newblock Internal thermodynamic variables and the failure of microcracked
  materials.
\newblock {\em Journal of Non-Equilibrium Thermodynamics}, 26(2):167--189,
  2001.

\bibitem{VanVas01p}
P.~V\'an and B.~V\'as\'arhelyi.
\newblock Second {L}aw of thermodynamics and the failure of rock materials.
\newblock In J.~P.~Tinucci D.~Elsworth and K.~A. Heasley, editors, {\em Rock
  Mechanics in the National Interest V1}, pages 767--773, Lisse-Abingdon-
  Exton(PA)-Tokyo, 2001. Balkema Publishers.
\newblock Proceedings of the 9th North American Rock Mechanics Symposium,
  Washington, USA, 2001.

\bibitem{Mir00a}
A.~Miranville.
\newblock Some generalizations of the {C}ahn-{H}illiard equation.
\newblock {\em Asymptotic Analysis}, 22(3-4):235--259, 2000.

\bibitem{BonMir01a}
Bonfoh A. and A.~Miranville.
\newblock On {C}ahn-{H}illiard-{G}urtin equations.
\newblock {\em Nonlinear Analysis - Theory Methods and Applications},
  47(5):3455--3466, 2001.

\bibitem{Aif80a1}
E.~C. Aifantes.
\newblock On the problem of diffusion in solids.
\newblock {\em Acta Mechanica}, 37:256--296, 1980.

\bibitem{GuyKru66a1}
R.~A. Guyer and J.~A. Krumhansl.
\newblock Solution of the linearized phonon {B}oltzmann equation.
\newblock {\em Physical Review}, 148(2):766--778, 1966.

\bibitem{GuyKru66a2}
R.~A. Guyer and J.~A. Krumhansl.
\newblock Thermal conductivity, second sound and phonon hydrodynamic phenomena
  in nonmetallic crystals.
\newblock {\em Physical Review}, 148(2):778--788, 1966.

\bibitem{PipRiv59a}
A.~C. Pipkin and R.~S. Rivlin.
\newblock The formulation of constitutive equations in continuum physics 1.
\newblock {\em Archive for Rational Mechanics and Analysis}, 4:129--144, 1959.

\bibitem{Smi64a}
G.~F. Smith.
\newblock On isotropic integrity bases.
\newblock {\em Archive for Rational Mechanics and Analysis}, 17:282--292, 1964.

\bibitem{Enz74a}
C.~P. Enz.
\newblock Two-fluid hydrodynamics description of ordered systems.
\newblock {\em Reviews of Modern Physics}, 46(4):705--753, 1974.

\bibitem{Fek81a}
D.~Fekete.
\newblock A systematic application of {G}yarmati's {W}ave {T}heory of
  {T}hermodynamics to thermal waves in solids.
\newblock {\em Phys. Stat. Sol. (b)}, 105:161--174, 1981.

\bibitem{Bha82p}
D.~K. Bhattacharya.
\newblock {\em The wave approach of irreversible thermodynamics}.
\newblock Doctoral thesis, Hungarian Academy of Sciences, Budapest, December
  1982.

\bibitem{LebDau90a}
G.~Lebon and P.~C. Dauby.
\newblock Heat transport in dielectric crystals at low temperature: A
  variational formulation based on {E}xtended {I}rreversible {T}hermodynamics.
\newblock {\em Physical Review A}, 42(8):4710--4715, 1990.

\bibitem{Net93a}
R.~E. Nettleton.
\newblock Reciprocity and consistency in non-local {E}xtended {T}hermodynamics.
\newblock {\em Open Systems and Information Dynamics}, 2(1):41--47, 1993.

\bibitem{LebGre96a}
G.~Lebon and M.~Grmela.
\newblock Weakly nonlocal heat conduction in rigid solids.
\newblock {\em Physics Letters A}, 214:184--188, 1996.

\bibitem{LebAta98a}
G.~Lebon, D.~Jou, J.~Casas-V\'azquez, and W.~Muschik.
\newblock Weakly nonlocal and nonlinear heat transport in rigid solids.
\newblock {\em Journal of Non-Equilibrium Thermodynamics}, 23:176--191, 1998.

\bibitem{ValAta97a}
A.~Valentini, M.~Torrisi, and G.~Lebon.
\newblock Heat pulse propagation by second sound in dielectric crystals.
\newblock {\em Journal of Physics: {C}ondensed Matter}, 9:3117--3127, 1997.

\bibitem{Lib90b}
R.~L. Liboff.
\newblock {\em Kinetic Theory (Classical, Quantum, and Relativistic
  Descriptions}.
\newblock Prentice Hall, Englewood Cliffs, New Jersey, 1990.

\bibitem{Van02m}
P.~V\'an.
\newblock Weakly nonlocal continuum physics - the {G}inzburg-{L}andau equation.
\newblock cond-mat/0210402, 2002.

\bibitem{PenFif90a}
O.~Penrose and P.~C. Fife.
\newblock Thermodynamically consistent models of phase-field type for the
  kinetics of phase transitions.
\newblock {\em Physica D}, 43:44--62, 1990.

\bibitem{VanMus95a}
P.~V\'an and W.~Muschik.
\newblock Structure of variational principles in nonequilibrium thermodynamics.
\newblock {\em Physical Review E}, 5(4):3584--3590, 1995.

\bibitem{Far894a}
Gy. Farkas.
\newblock A {F}ourier-f\'ele mechanikai elv alkalmaz\'asai.
\newblock {\em Mathematikai \'es Term\'eszettudom\'anyi \'Ertes\'\i{}t\H{o}},
  12:457--472, 1894.
\newblock in Hungarian.

\bibitem{Haa18a}
A.~Haar.
\newblock A line\'aris egyenl{\H{o}}tlens\'egekr{\H{o}}l.
\newblock {\em Mathematikai \'es Term\'eszettudom\'anyi \'Ertes\'\i{}t\H{o}},
  36:279--296, 1918.
\newblock in Hungarian.

\bibitem{Far18a2}
Gy. Farkas.
\newblock A line\'aris egyenl{\H{o}}tlens\'eg k\"ovetkezm\'enyei.
\newblock {\em Mathematikai \'es Term\'eszettudom\'anyi \'Ertes\'\i{}t\H{o}},
  36:397--408, 1918.
\newblock in Hungarian.

\bibitem{HauKir02a}
R.~A. Hauser and N.~P. Kirchner.
\newblock A historical note on the entropy principle of {M}\"uller and {L}iu.
\newblock {\em Continuum Mechanics and Thermodynamics}, 14:223--226, 2002.
\newblock Lecture held on CIMRF'01, Berlin, 3-6 September, 2001.

\bibitem{Rud70b}
W.~Rudin.
\newblock {\em Real and Complex Analysis}.
\newblock McGraw- Hill, London-etc..., 1970.

\end{thebibliography}
\end{document}